\documentclass[aps,prx,nofootinbib,10pt,twocolumn]{revtex4-2}

\usepackage{amsmath,amssymb}
\usepackage{times,hyperref,graphicx}
\hypersetup{colorlinks=true,linkcolor=blue,citecolor=blue,filecolor=blue,urlcolor=blue}

\newcommand{\f}{{\mathrm{f}}}
\newcommand{\A}{{\mathrm{A}}}
\newcommand{\B}{{\mathrm{B}}}

\begin{document}

\title{Effective synchronization amid noise-induced chaos}

\author{Benjamin Sorkin}
\email{bs4171@princeton.edu}
\affiliation{Princeton Center for Theoretical Science, Princeton University, Princeton 08544, New Jersey, USA}

\author{Thomas A. Witten}
\email{t-witten@uchicago.edu}
\affiliation{Department of Physics and James Franck Institute, The University of Chicago, Chicago 60637, Illinois, USA}

\begin{abstract}
Two remote agents with synchronized clocks may use them to act in concert and communicate. This necessitates some means of creating and maintaining synchrony.  One method, not requiring any direct interaction between the agents, is to expose them to a common, environmental, stochastic forcing.  This ``noise-induced synchronization'' only occurs under sufficiently mild forcing; stronger forcing disrupts synchronization. We investigate the regime of strong noise, where the clocks' relative phases evolve chaotically. Using a simple realization of disruptive noise, we demonstrate effective synchronization. First, although the relative phases of the two clocks varied erratically, we confirm that they became statistically independent of initial conditions and hence equivalent after a well-defined timescale. Second, we show that an agent can estimate an effective phase that closely agrees with the other's phase. Thus, synchronization is practically attainable beyond the regime of conventional noise-induced synchronization. 
\end{abstract}

\maketitle

\section{Introduction}\label{sec:intro}

Living systems and man-made machines often depend on the cooperation amongst independent agents. This cooperation is often attained using chemical~\cite{DunlapCELL99,RustPNAS21,BanfalviBOOK}, electrical~\cite{BOOK:spikes1997,MainenSCI1995,SteveninckSCI97,PennPNAS2016,NemenmanPNAS17,NemenmanTN18}, or mechanical~\cite{TanREV18} periodic processes (``clocks''), operating in each agent.  agents with synchronized clocks can exhibit emergent collective phenomena and are able to communicate~\cite{SYNCbook,BOOK:sync2003,SumpterRoyal06,ArenasPHYSREP008}.  Since independent clocks inevitably lose synchronization over time, some means are needed in order to achieve and maintain synchronization. 

One such means is noise-induced synchronization, occurring when identical noninteracting oscillators are subjected to a common, nonlinear, stochastic perturbation from their environment~\cite{BOOK:sync2003,Pikovsky1984,PikovskyPRL1997,JensenPRE98,TeramaePRL2004,GoldobinPRE2005,NakaoPRL2007,NagaiPRE2009,NakaoPRE2005,EatonPRE2016,SongPRE2022}. It is distinct from synchronization due to interactions between the clocks and from phase locking unto an external clock. Instead, synchronization is enabled via the clocks' shared ambient noise. As such, noise-induced synchronization is the only means of maintaining synchronization between noninteracting clocks in a generic environment. When the noise is sufficiently mild, it can bring the two oscillators into synchrony over time. This has been well characterized experimentally in a range of physical contexts, such as electric circuits~\cite{NagaiPRE2009}, quantum optics signaling~\cite{YamamotoOE07}, superconducting qubit chains~\cite{LutzPRL22,LutzNC25}, and gene regulatory networks~\cite{ZhouPRL05}.
Synchronization can enable time-based communication between a sender and a receiver. If the sender emits an impulse when its clock is at a particular phase, the receiver can infer this phase by noting its own clock's phase at the moment of the impulse. 

There is no guarantee that two oscillators will synchronize under a given noise. In general, synchronization requires the strength of the noise to lie below a threshold~\cite{BOOK:sync2003}.  Beyond this threshold, the relative phase of the two oscillators devolves into chaos~\cite{TeramaePRL2004,SongPRE2022}.  In the latter case, synchronization is lost and with it the means of time-based communication outlined above.  Nevertheless, we show below that a form of synchronization persists in the statistical properties of these erratically evolving phases.  It permits each agent to define an effective phase that agrees with that of other agents to a quantitatively-predictable extent, thus enabling them to behave as though synchronized. We will refer to this property as effective synchronization.

We demonstrate this behavior using a simplified type of noise, consisting of randomly timed impulses.  For example, this may represent stochastically timed electric shocks delivered to a collection of firing neurons or injections of a chemical affecting the cell cycle~\cite{BanfalviBOOK} into a cell culture. 
We study the effect of these perturbations on a distribution of phases. Successive impulses create a sequence of phase distributions that in general depend on the initial distribution and the timings of the impulses. In below-threshold noise-induced synchronization, these distributions become narrow and unimodal, concentrated at a single phase.  For noises above the threshold, the distributions remain spread over the entire phase circle~\cite{SongPRE2022}.  

The above-threshold effective synchronization we report here arises by virtue of two properties: First, though the distributions change markedly with each impulse, we find that they become independent of the initial state after a finite and predictable number of impulses.  This means that two initially different oscillator ensembles become statistically equivalent when compared at a given instant~\cite{BuzziCMP99,GuptaMRL13,FroylandNONLIN14,FroylandERG19}. Second, the distributions can be tightly bunched into a few sharp peaks.  Indeed, when the threshold of synchronization is approached from above, the distributions show a high degree of order, since their average entropy becomes arbitrarily negative~\cite{SongPRE2022}.

To demonstrate this effect, we numerically follow two different initial phase distributions for two agents $\A$ and $\B$. 
We find that the estimated Kullback-Leibler divergence among the two distributions, measuring their statistical difference, decreases with more kicks to zero after a characteristic kick number denoted $K_\mathrm{m}$. Thus, the phase distributions are determined from the most recent $K_\mathrm{m}$ kicks.
We propose an effective phase $\varphi_\f$ for a given distribution that each agent may estimate independently. We find that the effective phases $\varphi_\f$ of each agent $\A$ and $\B$ closely agree.

In Sec.~\ref{sec:prel}, we define a simplistic phase-reduction~\cite{NakaoPRE2005} dynamics and our statistical sampling methods.  In Sec.~\ref{sec:conv}, we demonstrate the quantitative convergence of different initial distributions to a single one. In Sec.~\ref{sec:sync} we give an explicit example of a workable $\varphi_\f[q(\varphi)]$ and show the strong agreement attainable between the $\varphi_\f$'s of agents $\A$ and $\B$. In Sec.~\ref{sec:disc}, we note the limitations of our study and argue for the generality and usefulness of this effective stochastic synchronization mechanism.

\section{Preliminaries}\label{sec:prel}

\subsection{Impulsive perturbations as phase maps}\label{sec:prelI}

Noise-induced synchronization occurs for independent agents with identical clocks. For our purposes, a clock is a nonlinear dynamical system orbiting a stable limit cycle. To illustrate stochastic synchronization, we specialize to a class of noise consisting of discrete, identical impulses (kicks) occurring at random times. The kicks are such that the oscillator remains within the basin of attraction of the limit cycle, whenever it occurs.  

The effect of a kick on the clock can be compactly expressed via a phase-reduction prescription~\cite{NakaoPRE2005}:
The oscillator traces a periodic loop in its dynamical manifold with a period $T$.  We may label the points on this loop by a ``phase position'' $\varphi$ ranging from $0$ to $1$, with a designated point assigned as the phase origin $\varphi\equiv0$. Following Refs.~\cite{TeramaePRL2004,NakaoPRE2005}, we label other points to have phase values proportional to the time $\Delta t$ required for the oscillator to reach them from the phase origin: $\varphi\equiv\Delta t/T$.\footnote{Unlike Refs.~\cite{TeramaePRL2004,NakaoPRE2005}, we do not include degrading noise, different for each oscillator. This eliminates the need to treat random excursions from the limit cycle via additional ``radial" coordinates. We discuss the effect of degrading noise in Sec.~\ref{sec:disc}.} (This construction is equivalent to finding the action-angle variables in Hamiltonian mechanics~\cite{ArnoldBOOK}.) Thus, the phase of an oscillator that started at $\varphi^{(0)}$ advances at a constant speed, $\varphi^{(0)}+t/T$, where $\mathrm{mod}\,1$ is implied in any algebraic operation over the phases.

\begin{figure}
    \centering
    \includegraphics[width=0.99\linewidth]{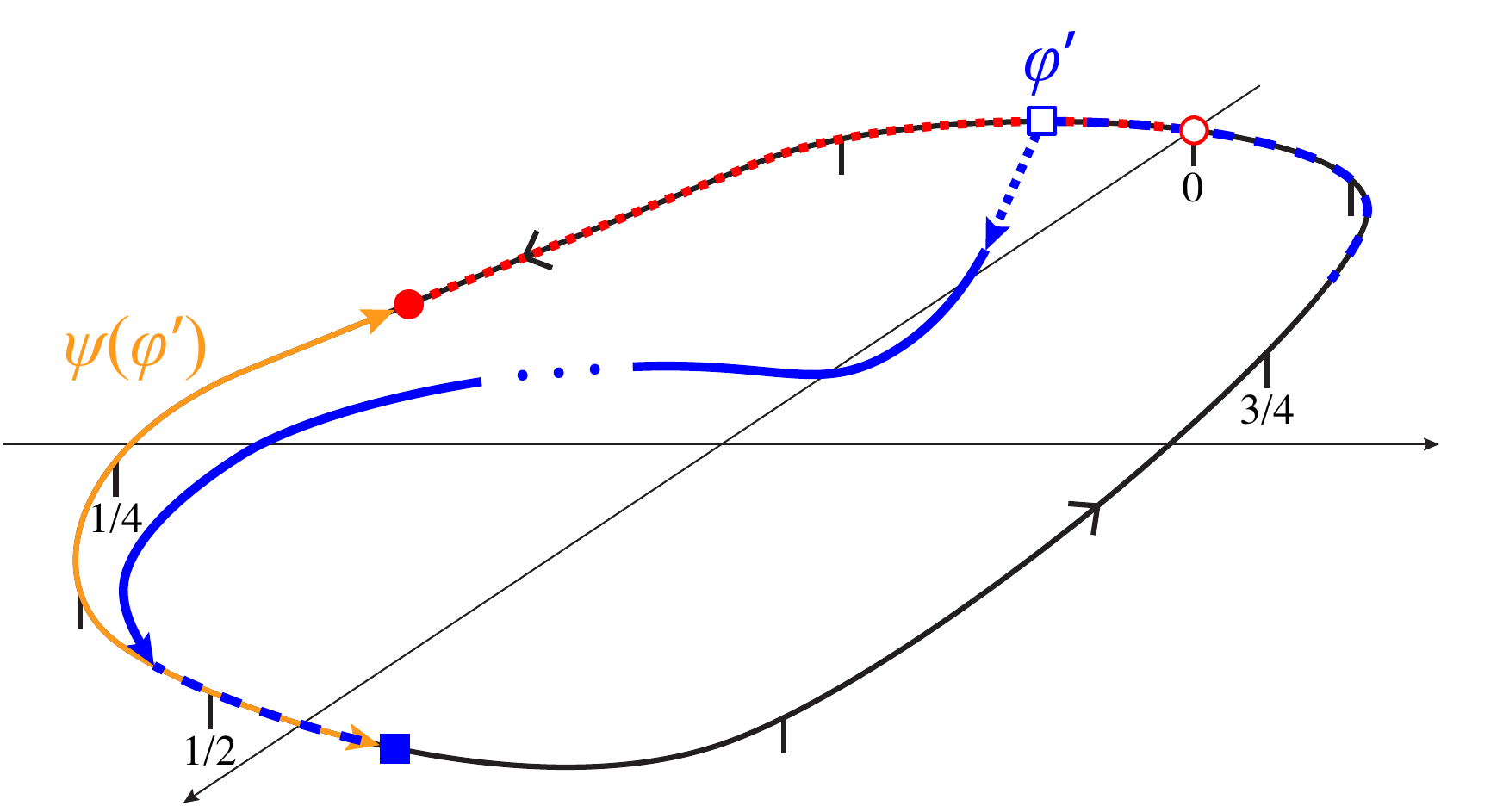}
    \caption{Illustration of a phase map $\psi(\varphi)$ determination for a given limit cycle.  A toy dynamical system is considered whose limit cycle is depicted by the closed black line in some dynamical phase space. Phase-position labels $\varphi$ along the loop are marked for the assigned phase origin $\varphi = 0$ and for phase positions displaced by $1/4$, $1/2$, and $3/4$ cycle from this origin. An open square marks the phase position of the oscillator at a moment of a kick, when its phase position was $\varphi'$. The subsequent trajectory of this oscillator is sketched in blue, showing its departure from the limit cycle and its eventual return to it. Before and after the kick, the oscillator gains phase at a constant speed (along the dashed trajectories). Also shown (in red) is the unperturbed trajectory of an oscillator which was at the phase origin at the moment of the kick. The position of the kicked trajectory at some arbitrary later time is marked with a filled square. The position of the unkicked trajectory at this same moment is marked with a filled circle.  Since both trajectories advance around the loop at the same rate, their phase difference, indicated in orange, is independent of time.  This difference for the arbitrary kick position $\varphi'$ is defined as the phase map, $\psi(\varphi')$.}
    \label{fig:illust}
\end{figure}

Every kick to an oscillator creates a fixed long-term shift in its time-dependent phase compared to the state of the oscillator if there had been no kick. This shift depends on the oscillator's phase position $\varphi'$ at the moment of the kick.  All effects described below result from the rearrangement of the oscillator positions due to this position-dependent shift. For impulsive kicks, it suffices to know the shift of an oscillator kicked at $\varphi'$, relative to the phase of an unkicked oscillator that was at the phase origin at the moment of the kick. This shifted phase, denoted $\psi$, is illustrated in Fig.~\ref{fig:illust}. Since all phase points move around the orbit at the same constant speed, the phase difference $\psi$ for a given oscillator remains constant until the next kick. Evidently, if the kick has a negligible effect, the phase of an oscillator\,---\,initially equal to $\varphi$\,---\,remains so, and thus $\psi(\varphi) = \varphi$. Otherwise, the ``phase map’’ function $\psi(\varphi)$ encodes the complete information about the long-term effect of the perturbation. 
 
A unique phase map $\psi(\varphi)$ can be determined for a given pairing of a dynamical system and a specified kick, as illustrated in Appendix~\ref{sec:example_phase_map}. In biological contexts, $\psi(\varphi)-\varphi$ is often called the phase-response curve, quantifying the response of circadian clocks~\cite{RustPNAS21,KhalsaJPHYSIOL03} and neural networks~\cite{PhaseResponseBOOK} to environmental changes. In these cases, the clock may be a living organism or periodically firing neurons, and the perturbations can be sporadic light impulses or shot noise from sensory neurons.

We study the case where this deterministic perturbation $\psi(\varphi)$ is applied at random times.\footnote{We assume that the waiting times in between kicks are sufficiently long so the oscillators have relaxed back to the limit cycle. Thus to arbitrary accuracy, the complete physical state of each oscillator may be specified by its phase along this cycle. For the case of possibly insufficiently long times, see Ref.~\cite{Hata:2010fk}.}
For simplicity, we may capture the effect of a $k$'th kick by observing the oscillators at an integer number of cycles after the latest kick. The unkicked position of the phase origin has continued its periodic motion and thus returns to the phase origin at such a moment. Likewise, the same phase is observed for the kicked oscillator after each completed cycle; we denote this phase as $\varphi^{(k)}$. Thus, the phase-map shift $\psi(\varphi')$ from the unkicked phase origin is simply $\varphi^{(k)}$. To complete the iteration, we must express the phase position at the moment of the $k$th kick $\varphi'$ in terms of the previous kick’s phase $\varphi^{(k-1)}$. During the random waiting time between the $k-1$'st and the $k$'th kick, the oscillator has gained a phase that we denote by $\beta^{(k)}$. Thus, $\varphi' = \varphi^{(k-1)} + \beta^{(k)}$.  Combining, we infer the effect of a single iteration,
\begin{equation}
\varphi^{(k)} = \psi(\varphi^{(k-1)} + \beta^{(k)}).\label{eq:dynamics}
\end{equation}
Immediately after a kick, this formula predicts the net phase shift that will be incurred after the oscillator has relaxed to its limit cycle. Due to the $\mathrm{mod}\,1$ constraint, only the fractional part of the waiting time in between kicks matters. For simplicity, we assume that $\{\beta^{(1)},\beta^{(2)},\ldots\}$ are identical independent uniformly-distributed random phases, $\beta^{(k)}\in[0,1)$. Equation~\eqref{eq:dynamics}, therefore, amounts to a Markov dynamics in the discrete ``time'' $k$, controlled by the phase map $\psi(\varphi)$ and the drawn $\beta^{(k)}$'s. Below, we compare the fates of two ensembles of oscillators exposed to the same sequence of $\beta^{(k)}$'s.\footnote{We consider synchronization of distinct initial phase distributions under the influence of a shared forcing, as opposed to identical initial phase distribution evolving under different forcings. We average over $\{\beta^{(k)}\}$'s only after computing differences between the two agents obtained with fixed $\{\beta^{(k)}\}$'s.}

Our aim in this paper is to examine generic behaviors of smooth phase maps. Accordingly, we consider simple cubic maps of the form
 \begin{equation}
    \psi(\varphi)=\varphi+A\varphi(1/2-\varphi)(1-\varphi),\label{eq:phase_map}  
\end{equation}
where the ``gain parameter" $A$ determines  the strength of the perturbations; when $A=0$, $\psi(\varphi) = \varphi$, so the kick has no effect. For completeness, we have repeated the paper's quantitative analyses also for the quintic phase map, $\psi(\varphi)=\varphi+A'\varphi(1-\varphi)(\Psi_1-\varphi)(\Psi_2-\varphi)(C-\varphi)$, where $C=(1-\Psi_1)(1-\Psi_2)/[\Psi_1\Psi_2+(1-\Psi_1)(1-\Psi_2)]$ is such that $(d\psi/d\varphi)|_{\varphi=1}=(d\psi/d\varphi)|_{\varphi=0}$. We will show the results for the cubic phase maps; we observed that the same effective synchronization occurs for the quintic ones. 

\subsection{Lyapunov exponent and synchronization}

With maps of this form, strong perturbations will typically be characterized by an increased gain parameter $A$. Once perturbed, the quantity $|d\psi/d\varphi|$ measures the degree of spreading of a local interval around $\varphi$. This ``spreading factor" therefore is closely related to the (in)ability of the phase map to synchronize adjacent phases. Hence, a measure of  effective strength of the kicks can be formulated by averaging this spreading factor.  Accordingly, we define $\Lambda$\,---\,the average of the log-spreading factor,\footnote{The integration measure is uniform since the waiting times are distributed uniformly $\beta\in[0,1)$, so every oscillator is likely to experience a kick with equal probability everywhere along the phase circle.}
\begin{equation}
    \Lambda=\int_0^{1}d\varphi\ln\left|\frac{d\psi}{d\varphi}\right|.\label{eq:lyapunov_exp}
\end{equation}
This is the average Lyapunov exponent~\cite{NakaoPRL2007,SongPRE2022} of the effective discrete-time dynamics in Eq.~\eqref{eq:dynamics}. 

Phase maps with $\Lambda<0$ tend to condense the phases. This means that oscillators that were subjected to the external force at common random waiting times $\beta^{(k)}$ are guaranteed to synchronize~\cite{NakaoPRL2007,SongPRE2022}. On the other hand, $\Lambda>0$ suggests that neighboring oscillators tend to spread apart in phase over successive kicks. This means that oscillators subjected to such forcings are guaranteed to not asymptote to a synchronized state in general. Increasing the gain parameter $A$ ultimately causes $\Lambda$ to cross the threshold from negative to positive.  This threshold distinguishes ``mild'' from ``strong'' perturbations.

\subsection{Probability distributions}

In order to analyze the stochastic states encountered with positive $\Lambda$, we consider the probability distribution $q^{(k)}(\varphi)$ arising from an initial $q^{(0)}(\varphi)$.  The evolution of this $q(\varphi)$ can be directly inferred from that of individual phases $\varphi$ in Eq.~\eqref{eq:dynamics}.

Prior to the perturbation, the phase distribution remains unchanged up to rotation: $q^{(0)}(\varphi,t)=q^{(0)}(\varphi-t/T)$ (see illustration in Appendix~\ref{sec:instantaneous}). We apply the same forcing (phase map $\psi$) after waiting times $\{\beta^{(k)}\}$ common to all oscillators. After the $k$'th kick, the ensemble will adopt the distribution $q^{(k)}(\varphi)$ according to
\begin{equation}
    q^{(k)}(\varphi)=\int_0^{1}d\varphi' q^{(k-1)}(\varphi')\delta(\varphi-\psi(\varphi'+\beta^{(k)})).\label{eq:qphi_dynamics}
\end{equation}
For $\Lambda<0$, synchronization manifests in a very narrow, unimodal (``single-peaked'') $q^{(k)}(\varphi)$'s which continually become narrower on average. In contrast, for $\Lambda>0$ one finds that $q^{(k)}(\varphi)$ continues to change erratically with each kick, with no ultimate convergence to narrow distributions.

\begin{figure}
    \centering
    {\large$\Lambda=-0.141$}
    \includegraphics[width=0.32\linewidth]{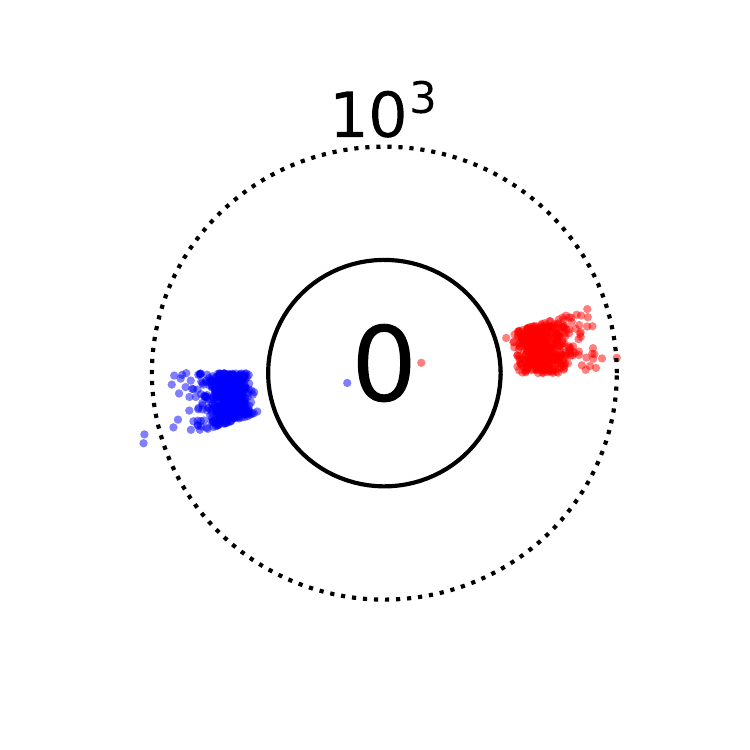}\includegraphics[width=0.32\linewidth]{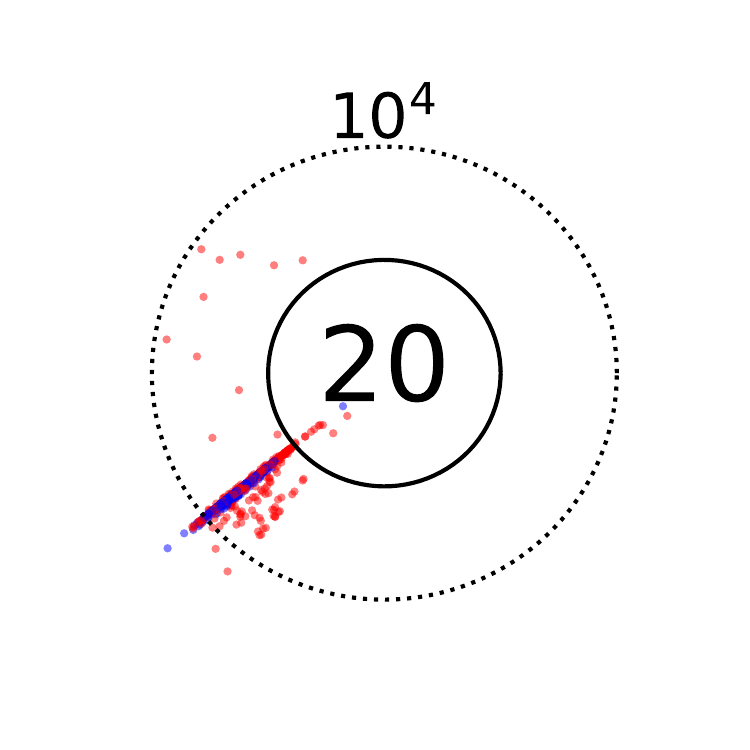}\includegraphics[width=0.32\linewidth]{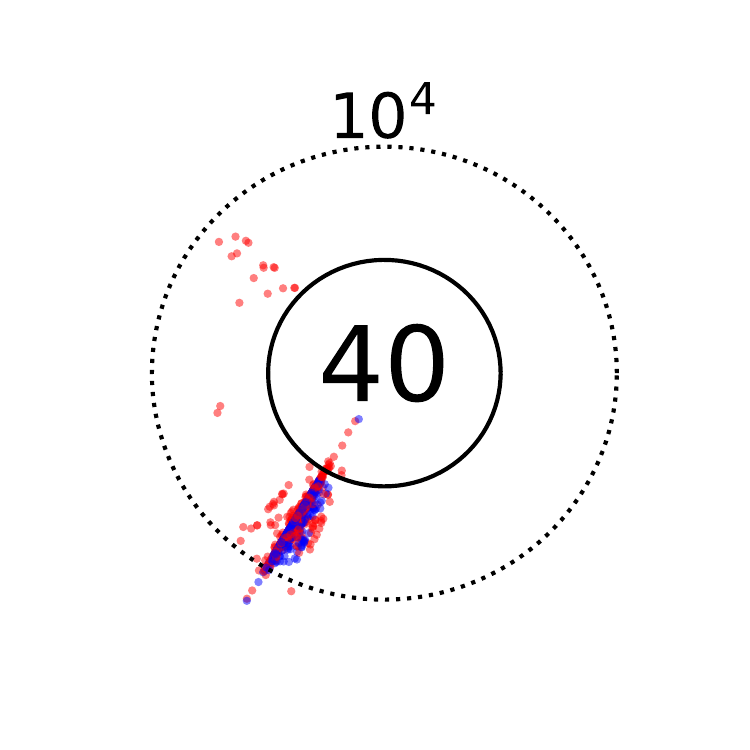}
    \includegraphics[width=0.32\linewidth]{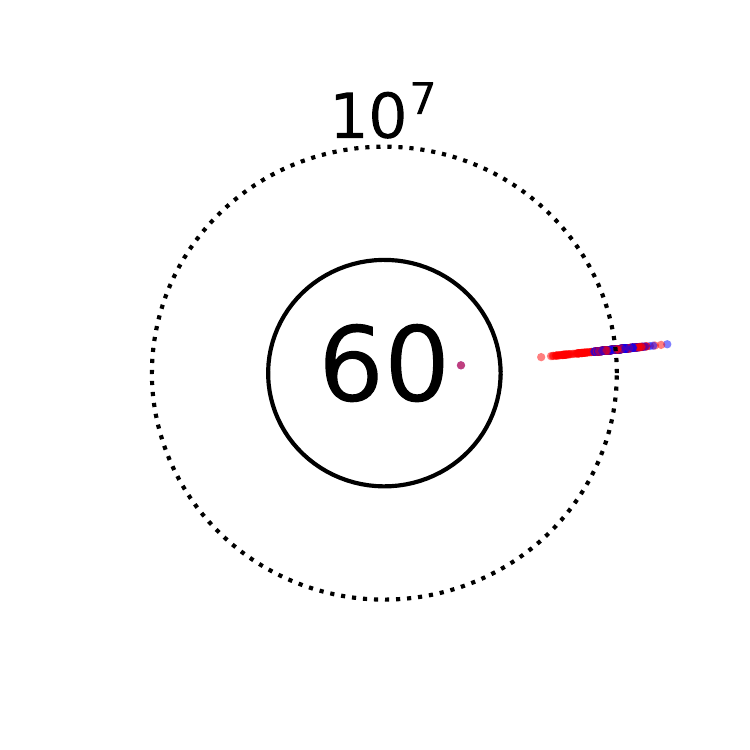}\includegraphics[width=0.32\linewidth]{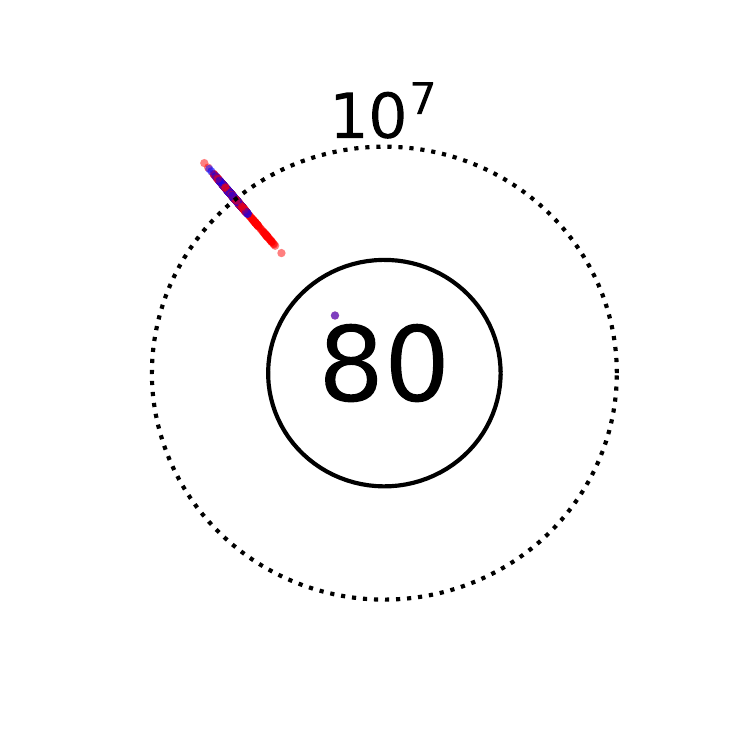}\includegraphics[width=0.32\linewidth]{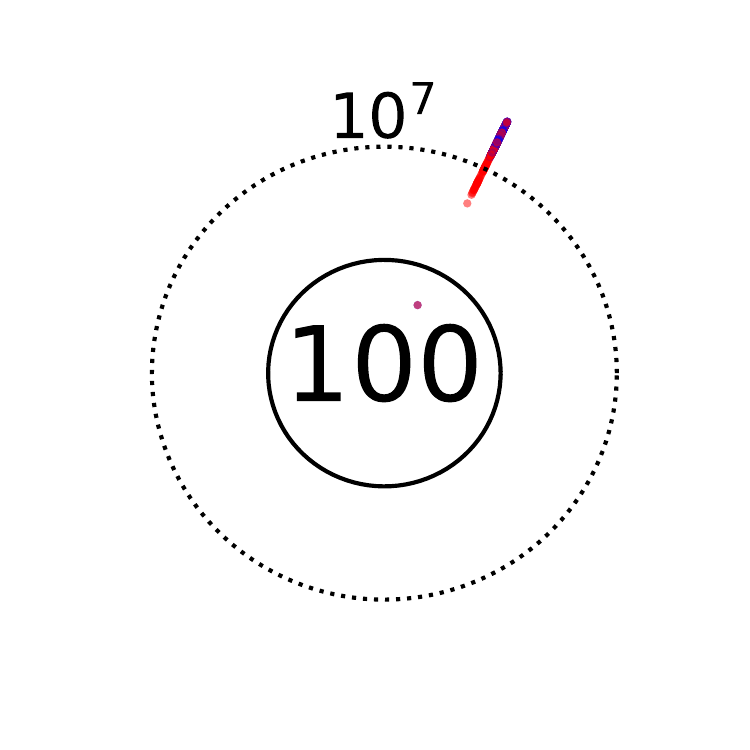}
    {\large$\Lambda=0.141$}
    \includegraphics[width=0.32\linewidth]{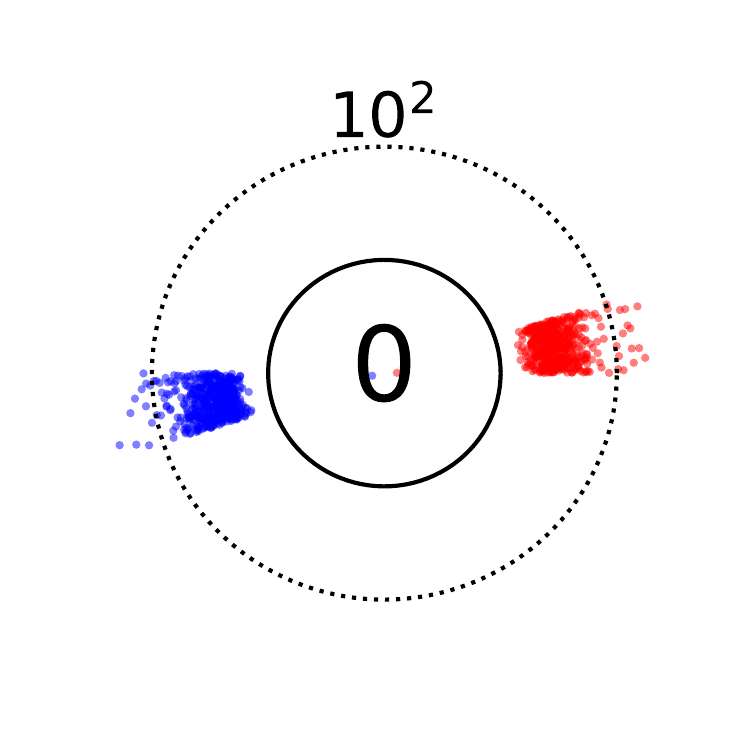}\includegraphics[width=0.32\linewidth]{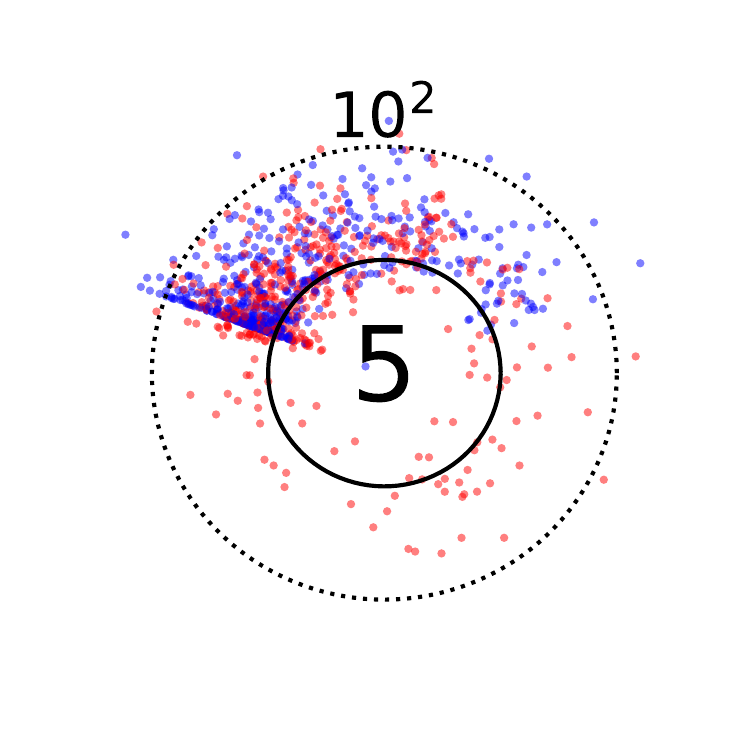}\includegraphics[width=0.32\linewidth]{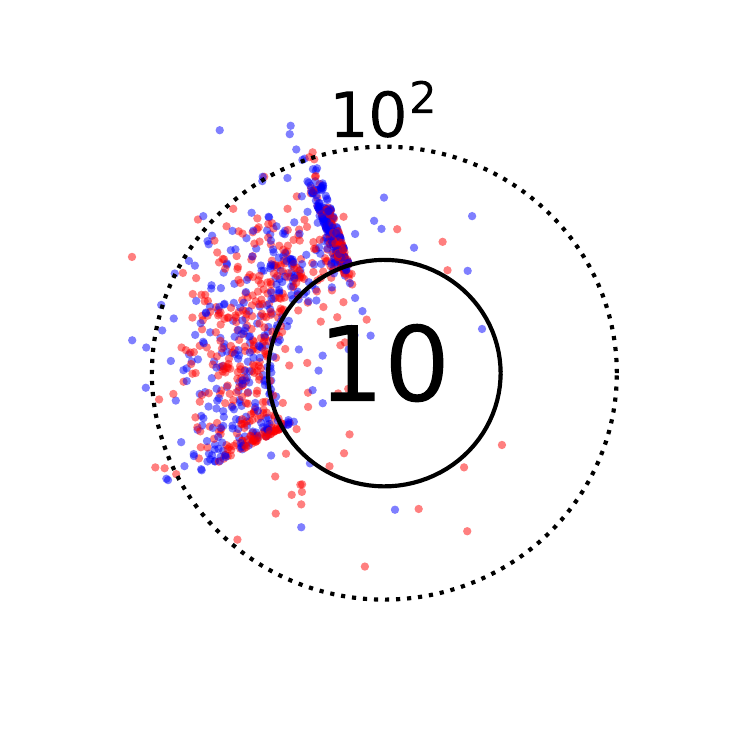}
    \includegraphics[width=0.32\linewidth]{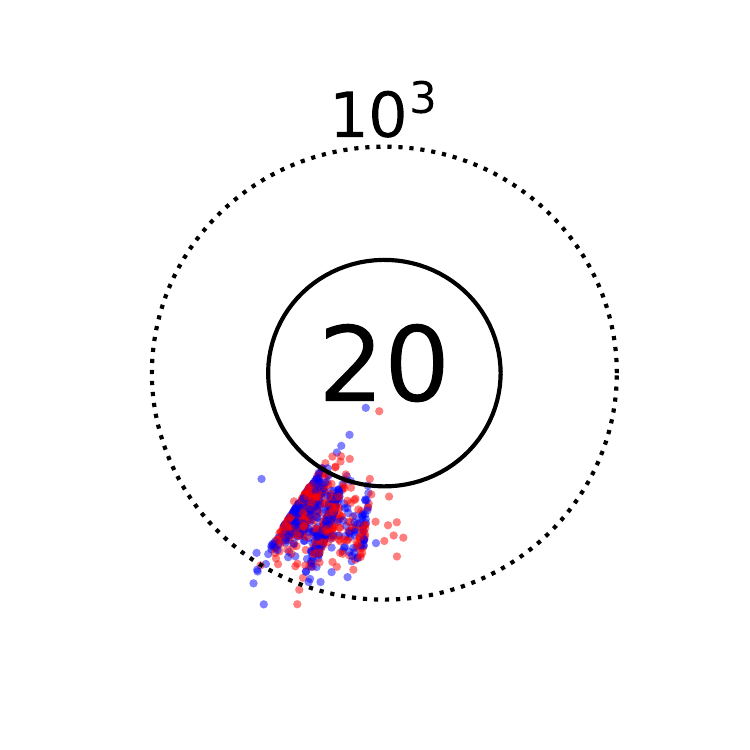}\includegraphics[width=0.32\linewidth]{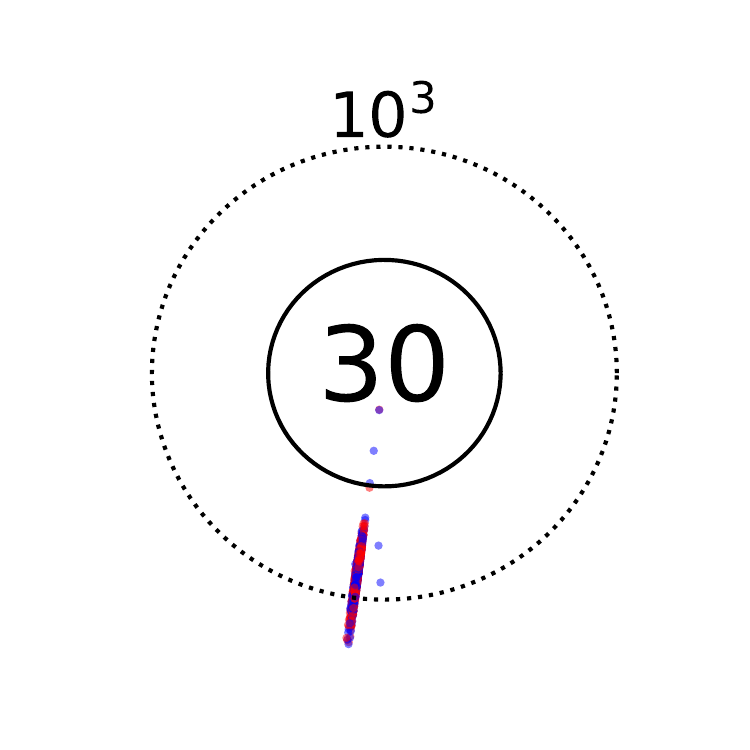}\includegraphics[width=0.32\linewidth]{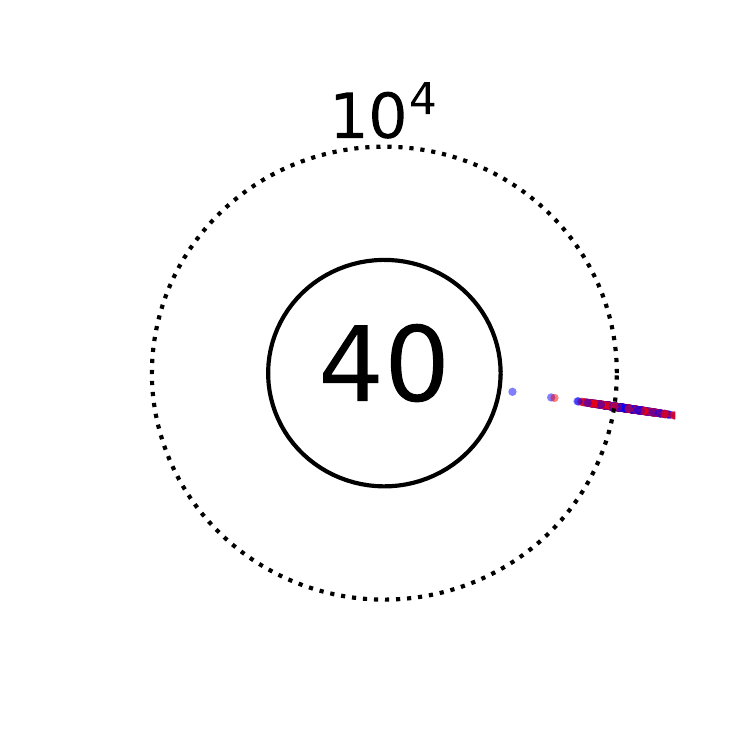}
    \includegraphics[width=0.32\linewidth]{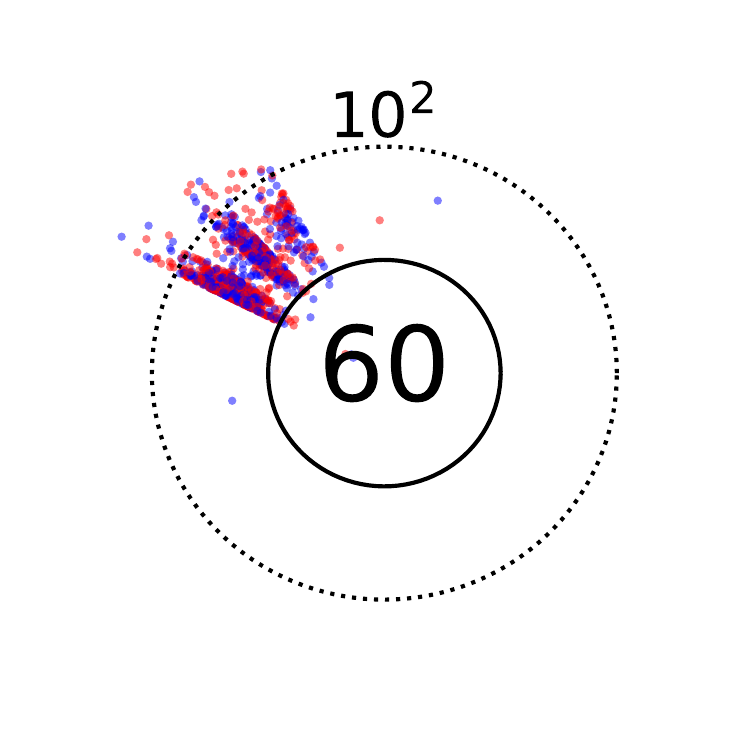}\includegraphics[width=0.32\linewidth]{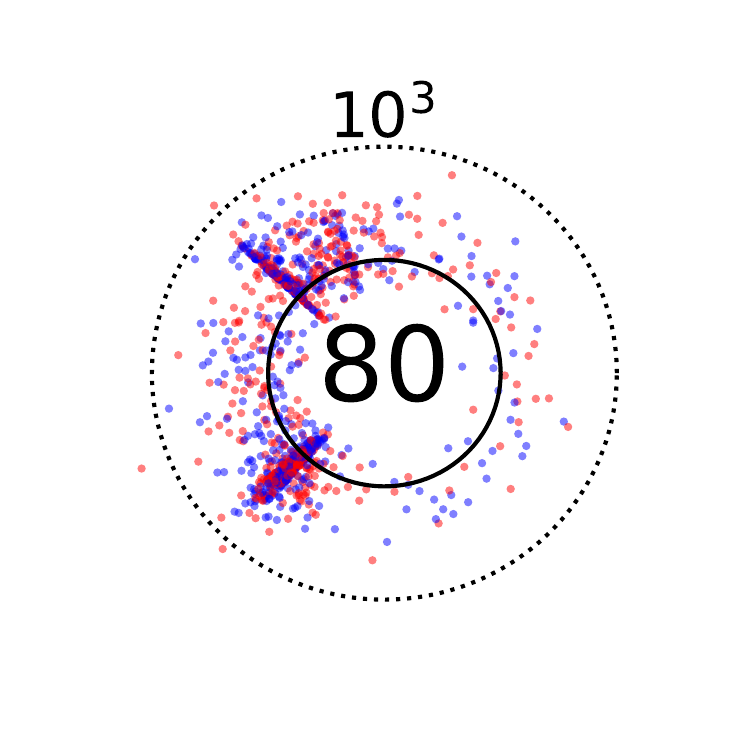}\includegraphics[width=0.32\linewidth]{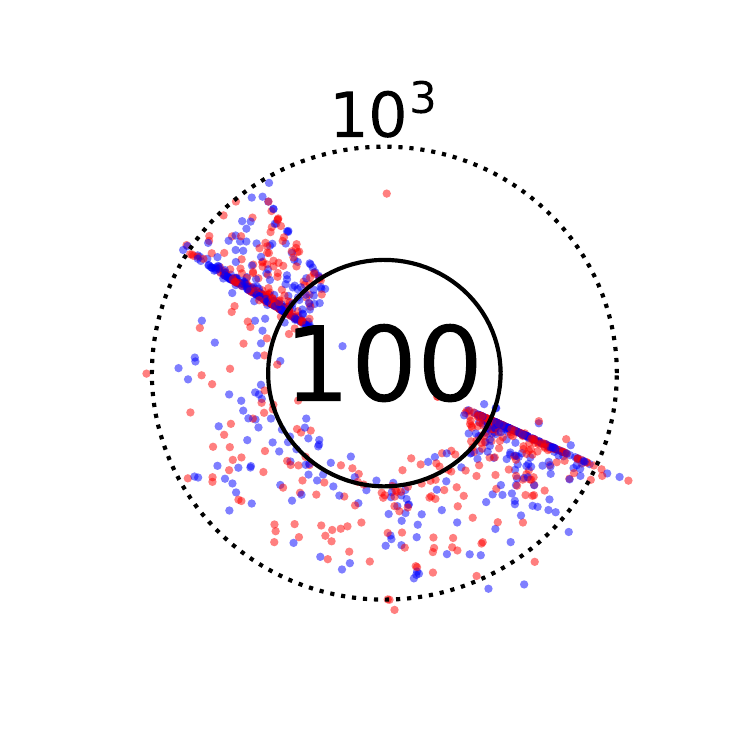}
    \caption{A sequence of two, initially-different phase distributions $q^{(k)}_\A(\varphi)$ (red) and $q^{(k)}_\B(\varphi)$ (blue) subjected to common noise in a typical realization of Eq.~\eqref{eq:dynamics} and Eq.~\eqref{eq:phase_map} with either $\Lambda=-0.141$ ($A=7.32$) or $\Lambda=0.141$ ($A=9.32$), as indicated. The initial distributions are uniform, $\varphi\in[0.0,0.05)$ and $\varphi\in[0.5,0.55)$, respectively. The distributions are shown on a circle so the periodicity $\,\mathrm{mod}\,1$ of the phase circle is apparent. The radial axis shows the distributions' values $q(\varphi)$ on a log scale, where $q(\varphi)=1$ for the inner full circle and as indicated for the outer dotted circles. The kick numbers $k$ are shown inside the inner circle. The $\Lambda<0$ dynamics are synchronizing, so their distributions become centered around the same phase value and their width decreases over time on average. For $\Lambda>0$, the distributions evolve erratically. However, remarkably, the red and blue points co-locate; they converge unto the same set of sharply-multimodal distributions under common noise\,---\,this property enables the effective synchronization demonstrated below. The distributions are sampled by tracking $N=500$ initial phases via the nearest-neighbor distances; see Eq.~\eqref{eq:est_q-phi}. The entropies of the two distributions and the KLDs among them are shown in Fig.~\ref{fig:LposS&KLD} for $\Lambda=0.141$ and Appendix~\ref{sec:convergenceL<0} for $\Lambda=-0.141$. A detailed view of the blue distribution at $k=100$ for $\Lambda=0.141$ is shown in Appendix~\ref{sec:misidentification}.}
    \label{fig:circ_q-phi}
\end{figure}

We now compare the experience of two independent agents, $\A$ and $\B$.  Each has a copy of the clock and a knowledge of its phase map $\psi(\varphi)$. The two clocks have arbitrary phases. As time goes by, the two agents' clocks experience a set of standard kicks at identical but random times. Each agent notes its oscillator's phase at the time of kick $k$ (this phase is $\varphi'$ in Fig.~\ref{fig:illust}) from which $\beta^{(k)}$ can be calculated as we outlined in Sec.~\ref{sec:prelI}. Knowing $\beta^{(k)}$, the agent can infer the phase shift incurred by any other clock $i$ whose phase was $\varphi_i^{(k-1)}$ before kick $k$, using Eq.~\eqref{eq:dynamics}.  Indeed, it can infer the phases of $N$ clocks whose initial phases $\varphi_i^{(0)}$ were chosen randomly.  It suffices to update the $i$'th phase at each successive kick. If agent $\A$ contains $N$ numerical registers that can each  store a number from $0$ to $1$, it can update each register with every kick and thus have $N$ samples of the phase after $k$ kicks.  As described below, the agent can then use these samples to obtain an estimate distribution $q_\A^{(k)}(\varphi)$ of Eq.~\eqref{eq:qphi_dynamics}. Proceeding similarly, agent $\B$ can obtain an independent estimate $q_\B^{(k)}$. In Fig.~\ref{fig:circ_q-phi}, we show the typical evolution of two distributions under common noise with $\Lambda<0$ and $\Lambda>0$.

In Sec.~\ref{sec:conv}, we find that after a finite number of kicks, the two distributions become operationally equivalent. This means that the same information is shared among the two agents, as both oscillators eventually follow the same statistics. To characterize this equivalence, we use the Kullback-Leibler divergence (KLD)~\cite{BOOk:info2006}\,---\,a natural means of quantifying the divergence of the distribution $q_\A(\varphi)$ from $q_\B(\varphi)$,
\begin{equation}
    D(\A\Vert\B)\equiv\int_0^1d\varphi q_\A(\varphi)\ln\frac{q_\A(\varphi)}{q_\B(\varphi)}.\label{eq:KLD}
\end{equation}
It quantifies the excess information stored in $q_\B$ when $q_\A$ is the presumed distribution. If the two distributions are the same, $q_\A(\varphi)= q_\B(\varphi)$, then $D(\A \Vert \B) = 0$. Otherwise, $D(\A \Vert \B) > 0$. We take $D(\A\Vert\B)\to 0$ to mean that the two distributions have converged and become equivalent.

Pursuing this evidence that all agents become statistically equivalent, we show in Sec.~\ref{sec:sync} that each agent may define an effective phase that agrees closely with that of the other agent. This is possible in spite of the agents' current phases differing widely. This agreement is possible owing to the special behavior of phase maps close to the $\Lambda = 0$ threshold.  Their typical $q(\varphi)$ distributions are strongly ordered, consisting of few narrow peaks, as will be seen below. To quantify the degree of order in a distribution we use the information entropy $S$~\cite{BOOk:info2006} defined in our context by 
\begin{equation}
    S\equiv-\int_0^{1} d\varphi\,q(\varphi)\ln q(\varphi).\label{eq:entropy}
\end{equation}
The highest possible value $S=0$ is achieved only for the uniform phase distribution, $q(\varphi)=1$, while sharp multimodal (``multi-peaked'') distributions will have a strongly-negative entropy. From a thermodynamic perspective, changes in the entropy are a lower bound on the work needed to create the given ordered distribution from a fully random (uniform) one.

The convolution of Eq.~\eqref{eq:qphi_dynamics} poses numerous issues: First, it cannot be carried out analytically for an arbitrary $q^{(k-1)}(\varphi)$. Second, each extremum in $\psi(\varphi)$, denoted $\varphi=\varphi_\mathrm{m}$, imparts an integrable singularity on $q^{(k)}(\varphi)$ of the form $\sim|\varphi-\psi(\varphi_\mathrm{m})|^{-1/2}$; these keep accumulating with each kick. 
Further, determining $q(\varphi)$ in practice requires sampling it with a finite number $N$ of samples.   Thus, we sample the chosen initial distribution $q^{(0)}(\varphi)$ by $N$ samples $\{ \varphi^{(0)}_1,  \varphi^{(0)}_2,  ...  \varphi^{(0)}_N \}$ drawn from $q^{(0)}(\varphi)$.   Each of these $N$ samples changes under a kick via the phase map, according to Eq.~\eqref{eq:dynamics}.  By repeatedly applying Eq.~\eqref{eq:dynamics}, we obtain the $\{ \varphi^{(k)}_1,  \varphi^{(k)}_2,  ...  \varphi^{(k)}_N \}$, corresponding to a sample of $q^{(k)}(\varphi)$.  We may then use standard tools to estimate $S^{(k)}$~\cite{WangIEEE09} and $D^{(k)}(A||B)$~\cite{VictorPRE2002} in terms of the $\{\varphi^{(k)}_n\}$, as described in Appendix~\ref{sec:KLDentEst}.  (The numerical uncertainty owing to discrete sampling in our simulations was of order $0.2$ for $S$ and $0.05$ for $D$ with $N=500$.) As we show in this work, the distributions of Fig.~\ref{fig:circ_q-phi}, obtained by such a discrete sampling, carry sufficient information about a synchronized phase notwithstanding the above singularities.

\section{Convergent distributions}\label{sec:conv}

The synchronization scheme proposed in Sec.~\ref{sec:sync} requires that $q_\A^{(k)}$ and $q_\B^{(k)}$ converge to the same distributions.  In this section, we demonstrate this convergence quantitatively via the decay of the KLD to zero for $\Lambda>0$ after a finite number of kicks.

\begin{figure*}
    \centering
    \includegraphics[width=0.49\linewidth]{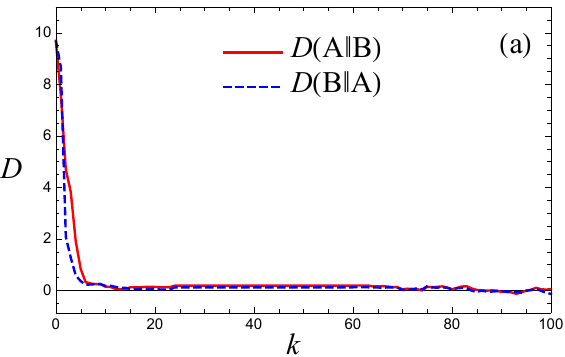}
    \put(-212,36){\includegraphics[width=0.37\linewidth]{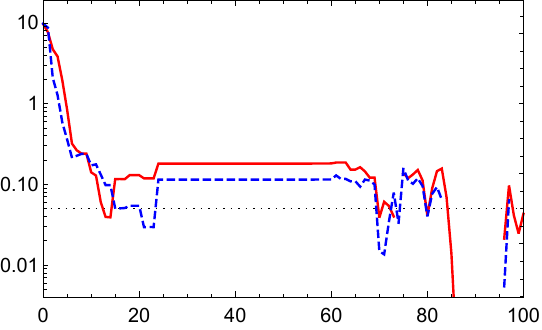}}
    \includegraphics[width=0.49\linewidth]{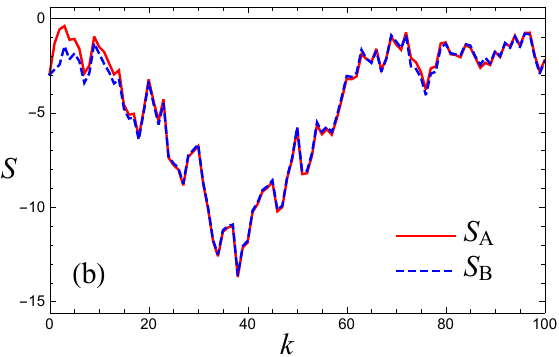}
    \caption{(a) Kullback-Leibler divergences (KLDs) $D(\A\Vert\B)$ and $D(\B\Vert\A)$ among two distributions and (b) entropies $S_\A$ and $S_\B$ of each distribution, as obtained for the waiting time realization $\{\beta^k\}$ of Fig.~\ref{fig:circ_q-phi} with $\Lambda=0.141$. Inset: The KLDs on a log-scale, where the numerical error in the KLD's estimation is indicates by the horizontal dotted line. The number of phase samples is $N=500$. Both initial distributions are of width $u=0.05$. The numerical errors in the estimation of the KLDs and entropies are $\Delta D=0.05$ and $\Delta S=0.2$, respectively.}
    \label{fig:LposS&KLD}
\end{figure*}

\begin{figure*}
    \centering
    \includegraphics[width=0.32\linewidth]{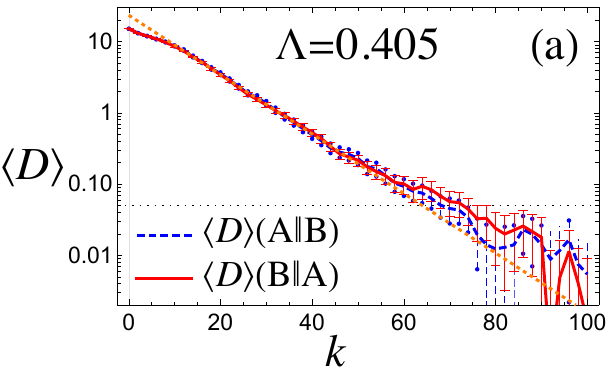}
    \includegraphics[width=0.32\linewidth]{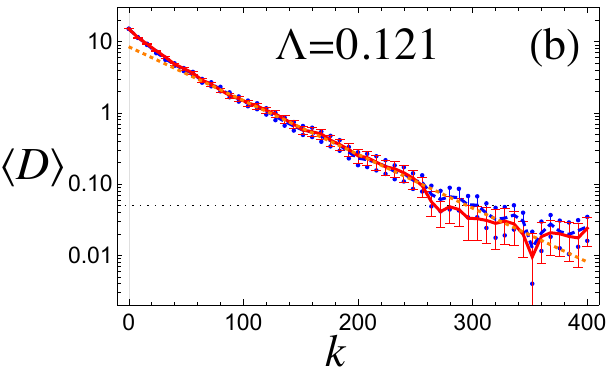}
    \includegraphics[width=0.32\linewidth]{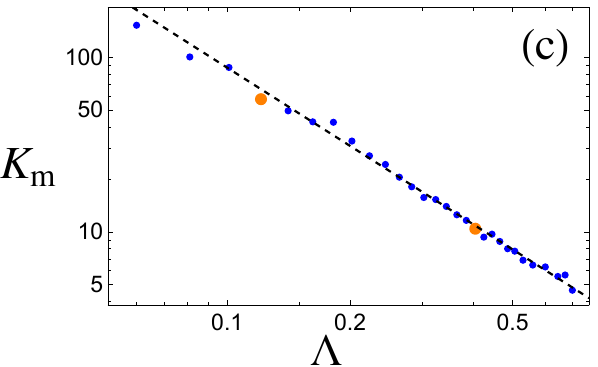}
    \caption{The average Kullback-Leibler divergence (KLDs) $D(\A\Vert\B)$ and $D(\B\Vert\A)$ among the two distributions, averaged over $500$ waiting-time realizations.  We used phase maps of Eq.~\eqref{eq:phase_map} with various values of $\Lambda$ obtained by varying the gain parameter $A$. We show the results for (a) $\Lambda=0.405$ and (b) $\Lambda=0.121$. The number of phase samples is $N=500$ in each ensemble, and they are initially uniform-distributed with width $u=10^{-4}$. Dashed orange lines depict the exponential fit from which we extract the mixing-kick number. (c) The mixing-kick number $K_\mathrm{m}$ versus Lyapunov exponent $\Lambda$. The two enlarged orange points correspond to $K_\mathrm{m}$ for the $\Lambda$'s shown in panels (a) and (b). $K_\mathrm{m}$ is identified as minus the inverse slope in the exponential decay of $\langle D\rangle$. The exponential regime begins after a transient where the distributions spread and overlap, and terminates when the KLD is comparable with its numerical estimation error, $\sim0.05$. Mixing occurs faster as $\Lambda$ increases, so the two agents typically converge earlier. The decay is consistent with the powerlaw $K_\mathrm{m}=2.82\Lambda^{-1.49}$.}
    \label{fig:tau_vs_lambda}
\end{figure*}

To this end, we simulate two oscillator ensembles, each consisting of $N=500$ phase samples. Their phase values are randomly and independently selected from different uniform distributions $\varphi^{(0)}_{\A n}\in[0,u)$ for the sender and $\varphi^{(0)}_{\B n}\in[0.5,0.5+u)$ for the receiver. By this, we study the extreme case of well-separated (non-overlapping) initial phase distributions.\footnote{The true KLDs should have started from $\infty$ as the two initial distributions are nonoverlapping. Instead, in Figs.~\ref{fig:LposS&KLD}(a) and~\ref{fig:tau_vs_lambda}(a,b) they begin at a large finite value, $D^{(0)}\sim {-\log u}$. This is an artifact of the numerical KLD estimate (Eq.~\eqref{eq:est_KLD}). The exponential decay due to the mixing we report commences past that short transient, at which point the two distributions have the same support $\varphi\in[0,1)$, wherein indeed the KLD is a reliable measure.} We subject the oscillators to Eq.~\eqref{eq:dynamics} for given Lyapunov exponents $\Lambda$ and initial widths $u$. All $2N$ phases evolve under the same $\psi(\varphi)$ and uniformly-distributed waiting times $\beta^{(k)}\in[0,1)$. At each kick, we estimate $S^{(k)}_\A$, $S^{(k)}_\B$, $D^{(k)}(\A\Vert \B)$, and $D^{(k)}(\B\Vert \A)$.\footnote{The KLD is not symmetric. However, since the initial distributions are offset by exactly half of the phase circle and $\beta\in[0,1)$, we expect both $D^{(k)}(\A\Vert\B)$ and $D^{(k)}(\B\Vert\A)$ to have the same statistics over many repetitions. Indeed their averages over realizations are equal in Fig.~\ref{fig:tau_vs_lambda}(a,b).}

For completeness, in Appendix~\ref{sec:convergenceL<0} we show how the synchronization for $\Lambda<0$ manifests in the entropy (see also Ref.~\cite{SongPRE2022}) and the KLD. For $\Lambda>0$, the two distributions $q_\A^{(k)}(\varphi)$ and $q_\B^{(k)}(\varphi)$ change erratically between consecutive kicks, \textit{e.g.}, as peaks are formed at phase values in the vicinity of the phase map's extrema, and `smeared' away from them (see Fig.~\ref{fig:circ_q-phi} and Appendix~\ref{sec:est_q-phi}). Accordingly, in Fig.~\ref{fig:LposS&KLD}(b), we see that the entropy also changes chaotically~\cite{SongPRE2022}. At the same time, both numerical KLDs decay to zero as seen in Fig.~\ref{fig:LposS&KLD}(a) despite the distributions being different initially, indicating that the two ensembles eventually converge to the same $k$-dependent distributions.\footnote{The brief transient during $25\lesssim k\lesssim 65$, where the KLD seems to stay at a constant value is a result of the distributions being very narrow at these time steps (see Fig.~\ref{fig:circ_q-phi}), so they change widths together without affecting the KLD. We quantitatively explain this behavior in Appendix~\ref{sec:convergenceL<0}. When the distribution widen afterwards, the KLD decreases to below its numerical error cutoff.} Accordingly, the entropies of both distributions coincide at later times in Fig.~\ref{fig:LposS&KLD}(b). Note that a different realization of waiting times $\{\beta^{(k)}\}$ produces vastly different distributions (see Appendix~\ref{sec:uncommon_noise}), implying that common noise is really key to this effect. 

This ensemble convergence can be explained qualitatively as arising from independence of initial state. Consider the map from an initial phase $\varphi^{(0)}$ to the final phase $\varphi^{(k)}$ for a particular sequence of random and uniformly distributed waiting times $\{\beta^{(k)}\}$. This map comes from a statistically uniform sampling of the phase map $\psi(\varphi)$ and has an overall Lyapunov exponent for any $\varphi^{(0)}$ that approaches $k\Lambda$ as $k$ gets large. Since the phase circle is periodic and finite, nearby phases cannot be drifting apart as $\sim e^{k\Lambda}$ indefinitely. Instead, the non-monotonic phase map (Eq.~\eqref{eq:phase_map}) ``stretches and folds'' phase samples, such that mixing of $\varphi^{(0)}$s is possible. As a result, the $\varphi^{(k)}$ for a given $\varphi^{(0)}$ may be arbitrarily close to the $\varphi'^{(k)}$ for other $\varphi'^{(0)}$'s residing anywhere throughout the phase circle. Then, any uncertainty in an observed $\varphi^{(k)}$ would erase all knowledge of $\varphi^{(0)}$. The asymptotic convergence of phase distributions under stochastic maps has been rigorously proven for general classes of stochastic systems~\cite{BuzziCMP99,GuptaMRL13,FroylandNONLIN14,FroylandERG19,BOOK:RandAttract}, based on spreading-of-phases arguments akin to the above.

To demonstrate applicability, however, we must go a step further and show that for any given uncertainty in $\varphi^{(k)}$ there exists some well-defined and finite kick number $K_\mathrm{m}$ such that any dependence of $\varphi^{(k)}$ on $\varphi^{(0)}$ for $k>K_\mathrm{m}$ is undetectable~\cite{MarkovMixingBook}. This diminishing detectability would entail a KLD that approaches zero, implying that the two distributions differ only by this decaying uncertainty. We thus proceed to confirm numerically that the KLD in our examples indeed converges to zero after a systematically predictable mixing number of kicks.

\subsection{Convergence rate}

Fig.~\ref{fig:LposS&KLD} is a single realization of stochastic waiting times $\{\beta^{(k})\}$. To find a typical time $K_\mathrm{m}$ that the agents should wait until their distributions become similar, we simulate a larger sampling of KLDs. To observe as many reliable KLD values as possible prior to their decay below the numerical-error cutoff of $\Delta D=0.05$, we picked the two initial distributions to be of width $u=10^{-4}$, so they are far apart and very narrow. We simulate $500$ realizations of $D^{(k)}(\A\Vert\B)$ and $D^{(k)}(\B\Vert\A)$ dynamics, each repetition sampling the above initial distribution and picking $\{\beta^{(k)}\}$ anew. We plot, as an illustration, the average KLDs and their statistical uncertainty versus $k$ for $\Lambda=0.405$ in Fig.~\ref{fig:tau_vs_lambda}(a) and $\Lambda=0.121$ in Fig.~\ref{fig:tau_vs_lambda}(b).

For all $\Lambda$'s, we observe a predominant regime of an exponential decay with $k$, shown with a dashed orange lines in Fig.~\ref{fig:tau_vs_lambda}(a,b). We anticipate a decay following the mixing argument of the preceding section. This exponential decay is typically cleaner for higher $\Lambda$'s (compare $k\in(10,45)$ in Fig.~\ref{fig:tau_vs_lambda}(a) with $k\in(70,250)$ in Fig.~\ref{fig:tau_vs_lambda}(b)), and lasts until we reach the numerical error cutoff for the KLD estimator, $\simeq0.05$. 
Thus, for each $\Lambda$, we fit an exponential (as in Figs.~\ref{fig:tau_vs_lambda}(a,b)) to obtain a decay rate defined as $1/K_\mathrm{m}(\Lambda)$. We interpret these $K_\mathrm{m}$'s as the mixing-kick number. We verify for all $\Lambda$'s that it does not depend on the initial condition (by simulating $u=5\cdot10^{-4}$ as well) or the sample size (by simulating $N=100$, too); $K_\mathrm{m}(\Lambda)$ remained identical within $3\%$ for all $\Lambda$'s, indicating that this convergence kick number is a robust quantity intrinsic to the dynamical system studied.  

In Fig.~\ref{fig:tau_vs_lambda}(c) we plot the mixing-kick number $K_\mathrm{m}$ versus $\Lambda$. For the cubic map, we numerically find that $K_\mathrm{m}$ strongly depends on $\Lambda$, varying across two orders of magnitude in a manner consistent with a power law, $K_\mathrm{m}(\Lambda)\sim \Lambda^{-1.491\pm0.046}$. This implies that the more violent the phase map,\footnote{By more violent we mean a larger gain parameter $A$ of the cubic map, Eq.~\eqref{eq:phase_map}. For $\Lambda>0$, $\Lambda(A)$ is a monotonously increasing function for the cubic map; see, \textit{e.g.} Fig.~15.3 in Ref.~\cite{BOOK:sync2003}.} the faster the mixing in the considered dynamics. We caution that the exponential fit for the decay of the KLD with low $\Lambda$ is not perfect (specifically, for $\Lambda\lesssim0.2$), so their $K_\mathrm{m}$ should be taken with a grain of salt. 

We have no quantitative argument as to why $K_\mathrm{m}$ should diverge as $\Lambda \to 0^+$ as our findings suggest. Generally, the mixing property governing $K_\mathrm{m}$ is in principle different from the spreading property that governs $\Lambda$. Qualitatively, the spreading of phases seems to be the only effect enabling mixing, so the absence of spreading ($\Lambda\to0^+$) should imply the absence of mixing ($1/K_\mathrm{m}\to0$). From a different perspective, we point out in Sec.~\ref{sec:theor_impl} that $\Lambda=0$ marks the transition between a random point attractor and an extended random attractor~\cite{BOOK:RandAttract}. In any case, the exact functional form of $K_\mathrm{m}(\Lambda)$ is not central to what follows; we will use the fact that $K_\mathrm{m}$ diverges as $\Lambda\to0^+$ and that it is well-defined and finite otherwise as a qualitative guide.

To summarize, for $\Lambda>0$, the two phase statistics become identical after iterating sufficiently longer than the mixing-kick number $K_\mathrm{m}$. Thus, for the purpose of establishing synchronization, we will first make the reasonable assumption that both agents have received a number of common kicks exceeding an agreed threshold. In the next section we demonstrate this synchronization.

\section{Effective synchronization}\label{sec:sync}

In Sec.~\ref{sec:conv}, we found that after finitely many kicks, agents $\A$ and $\B$ effectively share $q^{(k)}(\varphi)$ in both the $\Lambda<0$ (synchronizing) and the $\Lambda>0$ (unsynchronizing) regimes. With $\Lambda<0$, the two agents agree at every moment (sufficiently-long after the most recent kick) on a specific phase to experimental accuracy and can use it to perform simultaneous actions or communicate. 
For $\Lambda>0$, while the resultant distributions do not asymptotically approach a sharp peak, the distributions of two agents become equivalent as numerically captured by the approach of the KLD among them to zero. The synchrony in the former and convergence in the latter remain with arbitrarily more kicks. Thus, any function of the sampled phases $\{ \varphi^{(k)}_1,\ldots,\varphi^{(k)}_N \} $ would be equal for both agents up to a statistical error. Below, we will consider a particular phase function we refer to as the ``fiducial phase'', denoted $\varphi^{(k)}_\f=\varphi_\f[ \varphi^{(k)}_1,\ldots,\varphi^{(k)}_N ]$. By phase function we mean that if $\{\varphi_{n}^{(k)}\}$ shift by $\Delta\varphi$, it must also shift by $\Delta\varphi$.

In order for agent $\A$ to act as though synchronized with agent $\B$, it must be able to infer agent $\B$'s fiducial phase $\varphi_{\f, \B}$ at any given moment, given its phases $\{\varphi_{\A n}^{(k)}\}$. If agent $\A$'s oscillator at that moment is $\varphi^*_\A$, its own fiducial phase $\varphi^*_{\f, \A}$ at that moment may be determined, since all phases advance at the same rate; see Appendix~\ref{sec:instantaneous}. Specifically, $\varphi^*_{\f, \A} - \varphi^*_\A = {\rm const} = \varphi^{(k)}_{\f, \A} - \varphi^{(k)}_\A$; the same applies to agent $\B$. Therefore, it remains to be shown that one may construct a function $\varphi^{(k)}_{\f}$ such that the estimated $\varphi^*_{\f, \A} =  \varphi^*_{\f, \B}$ to a good accuracy. We demonstrate below that this type of synchronization can indeed be achieved.

The uncertainty in $\varphi_\f$ must depend on the degree of nonuniformity of $q(\varphi)$; surely, no $\varphi_\f$ can be unambiguously determined if $q(\varphi)$ is completely uniform.  Further, uncertainty in measuring $q(\varphi)$ leads to further uncertainty in $\varphi_\f$.  Below we test the discrepancy between $\varphi_{\f,\A}$ and $\varphi_{\f,\B}$ using the finite sampling method employed above. Our proof-of-concept example below shows that with a fixed sample size $N$, it is indeed feasible to attain small discrepancies which decrease with bigger $N$.  These discrepancies depend strongly on the nonuniformity of the $q(\varphi)$ at hand, as measured by its entropy, Eq.~\eqref{eq:entropy}.  Thus, reliably small discrepancies require phase maps $\psi(\varphi)$ with small typical entropies.

Here we follow a simple strategy to arrive at a choice of fiducial phase, $\varphi_\f$, for a given $q(\varphi)$.  We seek a functional that is defined to high precision when the $q(\varphi)$ is strongly nonuniform and concentrated into narrow peaks (cf. Fig.~\ref{fig:circ_q-phi}).  If there is a single peak, an obvious choice for $\varphi_\f$ is simply the position of the maximum.  If there are two or more maxima, one generally dominates over the other in the sense that it has the greatest height and covers more probability mass. This suggests a choice of $\varphi_\f$ as the position of the highest peak. 

To determine a peak position, one needs a smooth estimate of $q(\varphi)$ from the sampled values $\{\varphi_1,\ldots,\varphi_N\}$. We obtain an estimate denoted $Q(\varphi, \sigma)$ smoothed to a ``bandwidth" $\sigma$ using the method of kernel-density estimation~\cite{bookSIGMA86,ParkSIGMA90,SheatherSIGMA91}, detailed in Appendix~\ref{sec:fiducial}. The reliability of the $\varphi_\f$ obtained depends on $\sigma$. Large $\sigma$ produces a $Q(\varphi;\sigma)$ with a few broad peaks whose positions have high uncertainty, whereas small $\sigma$ gives numerous statistically insignificant peaks; see Appendix~\ref{sec:smooth_q-phi}.\footnote{Finding an optimal bandwidth $\sigma$, particularly for highly-multimodal distributions, is an open question in the field~\cite{SilvermanSIGMAMULTI81,MammenSIGMAMULTI92,VincentSIGMAMULTI02,meszarosSIGMAMULTI24}.}

We observe that the multimodal distributions encountered for small positive $\Lambda$ have a dominant peak containing a substantial fraction of the $N$ samples (Fig.~\ref{fig:circ_q-phi}). In these narrowly-peaked distributions, the measured entropy $S$ (Eq.~\eqref{eq:entropy} and Eq.~\eqref{eq:est_entropy}) gives guidance about the appropriate $\sigma$: If all $\{\varphi_n\}$'s lie within an interval of width $w$, then $S=\ln w$ up to an additive constant~\cite{Entropy2018lower}. Now, if instead a substantial fraction of them lie within $w$, the entropy remains comparable to $\ln w$. Thus, the measured entropy for each $q(\varphi)$ gives a characteristic length indicative of the dominant peak's width. Thus, in what follows, we have simply used a bandwidth $\sigma$ equal to this $w$, \textit{i.e.}, $\sigma = e^S$, as the peak center will not be offset by more than its width. We explored other choices of $\sigma$ empirically to see their effect on the resulting uncertainties in $\varphi_\f$; see Appendixes~\ref{sec:smooth_q-phi} and~\ref{sec:sigma_vary}. Accordingly, our fiducial phase of choice is
\begin{equation}
    \varphi_\f=\mathrm{argmax}_{\varphi\in[0,1)} Q(\varphi;e^S).\label{eq:fiducial}
\end{equation}
Even with this crude choice, we are able to demonstrate a clear effective synchronization.

\subsection{Proof of effective synchronization}\label{sec:proof}

\begin{figure*}
    \centering
    \includegraphics[width=0.49\linewidth]{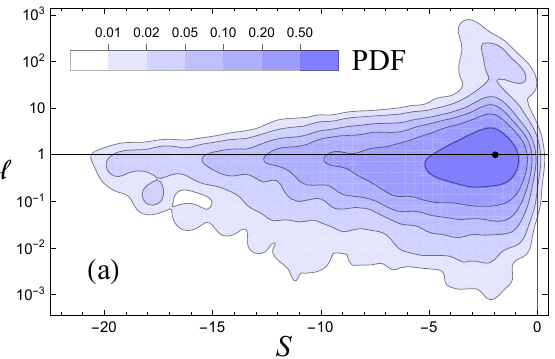}
    \includegraphics[width=0.49\linewidth]{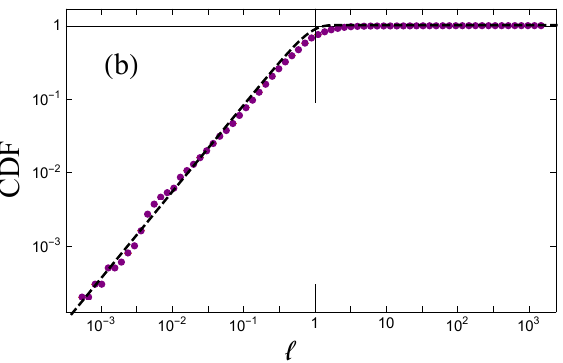}
    \put(-148,27){\includegraphics[width=0.28\linewidth]{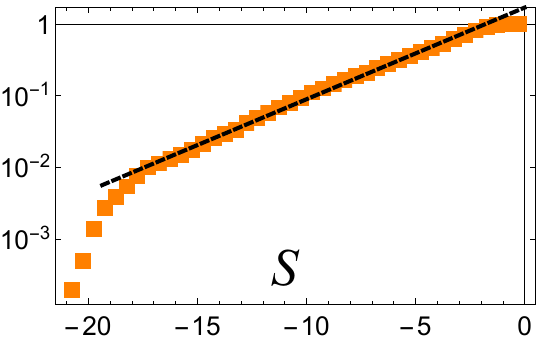}}
    \caption{(a) A density plot of the rescaled discrepancy, $\ell=\sqrt N|\Delta\varphi_\f|/e^S$ versus $S=(S_\A+S_\B)/2$. The shade of blue represents the indicated values of the joint probability density function (PDF) to obtain $S$ and $\log_{10}\ell$, relative to its maximum. It was computed with kernel-density estimation from the $k=10^4$ data points shown in Appendix~\ref{sec:sigma_vary} (Fig.~\ref{fig:deviation_vs_entropySM}(a)), using a Gaussian kernel of width $0.4$ in $S$ and $0.2$ in $\log_{10}\ell$. The maximal value is $\mathrm{PDF}(-1.94,-0.0052)=0.135$, whose position is indicated  by the black point. The puzzling lobe of probability at the upper right is addressed in Appendix~\ref{sec:sigma_vary}. (b) The marginal cumulative distribution function (CDF) of the discrepancy $\log_{10}\ell$, which was arranged into bins of size $0.25$. The dashed line is the error function of width unity; its close agreement with the data suggests that the discrepancy is predominantly normal-distributed with the expected scaling $|\Delta\varphi_\f|\sim e^{S}/\sqrt{N}$. Inset: The marginal CDF of the entropy $S$, which was arranged into bins of size $0.5$. The dashed linear line implies, up to the low- and high-entropy outliers, that the entropy is exponentially distributed~\cite{SongPRE2022}.}
    \label{fig:deviation_vs_entropy}
\end{figure*}

We now test the extent to which agent $\A$'s $\varphi_{\f,\A}^{(k)}$ and $\B$'s $\varphi_{\f,\B}^{(k)}$ agree. We use an extensive simulation of the phase map of Eq.~\eqref{eq:phase_map} with $\Lambda = 0.141$ and $N=500$. We extend the simulation depicted in Fig.~\ref{fig:circ_q-phi} to a large number of kicks, $k = 10^4$. For every $k$, using the procedure of Appendix~\ref{sec:fiducial} (and the definition in Eq.~\eqref{eq:fiducial}), we find $\varphi^{(k)}_{\f,\A}$ and $\varphi^{(k)}_{\f,\B}$.
To quantify the success of our scheme for identifying a common fiducial phase, we consider the statistics of the deviation $\Delta\varphi_\f=\varphi_{\f,\A}-\varphi_{\f,\B}$.
There are indeed $4,987$ positive and $5,013$ negative $\Delta\varphi_\f$ values, both exhibiting identical histograms (the KLD among the two $\ln(\pm\Delta\varphi_\f)$ distributions is $\mathcal{O}(10^{-3})$), so we only show $|\Delta\varphi_\f|$ henceforth. Likewise, in light of the convergence we observe in Sec.~\ref{sec:conv}, we shall only plot $S=(S_\A+S_\B)/2$, where $S_\A$ and $S_\B$ only differ by numerical entropy estimation error $\sim0.2$ for large $k\gg K_\mathrm{m}$.

Though $\Delta \varphi_\f$ varies widely between kicks, this variability stems from the erratically varying entropies of the distributions $q^{(k)}$ (Fig.~\ref{fig:LposS&KLD}(b)).  
In Fig.~\ref{fig:deviation_vs_entropy}(a), we plot the scaled deviation $\ell\equiv\sqrt{N}|\Delta\varphi_\f|/e^S$ versus $S$, where the color intensity depicts the joint probability density for $(S,\log_{10}\ell)$. We see that for all $S$'s, most $|\Delta \varphi_\f|$'s lie close to $e^S/\sqrt{N}$ (see the $\ell=1$ line). The deviations from the peak appear to fall off roughly as a Gaussian\,---\,we further resolve this in Fig.~\ref{fig:deviation_vs_entropy}(b). There, we plot the marginal cumulative distribution of the rescaled deviation $\ell$ irrespective of $S$, highlighting the abundance of small deviations as seen qualitatively in Fig.~\ref{fig:deviation_vs_entropy}(a). The inset shows the marginal cumulative distribution of the entropy $S$ irrespective of $\ell$. The raw data for Fig.~\ref{fig:deviation_vs_entropy}\,---\,the $10^4$ instances of deviations $|\Delta\varphi_\f|$ versus $S$\,---\,is shown in Appendix~\ref{sec:sigma_vary} (Fig.~\ref{fig:deviation_vs_entropySM}(a)).

Overall, both panels of Fig.~\ref{fig:deviation_vs_entropy} are most encouraging, as a common fiducial phase is established within uncertainty much less than $1$. Indeed the most abundant instances surrounding the black point ($\ell\simeq1.0$, $S\simeq-2.0$) have a phase discrepancy of order $|\Delta\varphi_\f|=1\cdot e^{-2}/\sqrt{500}=6.1\cdot10^{-3}$, which is a fraction of a percent synchronization accuracy. Further, much smaller deviations are abundant: From Fig.~\ref{fig:deviation_vs_entropy}(b), $10\%$ of the distributions produce reduced discrepancies smaller than $\ell=10^{-1}$, or $|\Delta \varphi_\f|=6.1\cdot10^{-4}$. The fitted error function ($\mathrm{erf}(x)=(2/\pi)\int_0^xdte^{-t^2}$) suggests that the discrepancy predominantly follows a normal distribution. This supports our proposed scaling $|\Delta\varphi_\f|\sim e^{S}/\sqrt{N}$, as indeed the predominant peak typically contributes most to the entropy, so the local width is comparable to $e^S$ and contains most of the phase samples ($\mathcal{O}(N)$), so the standard deviation in the estimation of the mean decays as $1/\sqrt{N}$.

Repeating the above for $\Lambda=0.101,0.141,\ldots,0.501$ reveals that the entropy is still exponentially distributed with a (negative) mean that increases with $\Lambda$, asymptoting to $0$ for large $\Lambda$, which is consistent with Ref.~\cite{SongPRE2022}. For all $\Lambda$'s tested and for $N=125,250,500$, all the corresponding plots fully overlap with Fig.~\ref{fig:deviation_vs_entropy}(b) (except for the statistically insignificant tails for $\ell<10^{-2.5}$). 

To conclude, this section has provided concrete evidence supporting the proposal made in Sec.~\ref{sec:intro}. Identical nonlinear oscillators exposed to identical noise can enable independent agents to act in concert and perform time-based communications as though their clocks were synchronized, even though they are not and their phases are only statistically determined.

\section{Discussion}\label{sec:disc}

In this paper, we showed that noise-induced synchronization can be extended beyond its recognized limits.  Our study demonstrated effective synchronization under noise-driven dynamics with positive Lyapunov exponents, which precludes conventional noise-induced synchronization. This effective synchronization arose from two key features of our dynamics. First, the statistical distribution of phases at a given moment becomes independent of the initial distribution for any $\beta^{(k)}$ realization after a well-defined finite typical mixing-kick number $K_\mathrm{m}$, as noted in Sec.~\ref{sec:conv}.  Thus, the distributions seen by two independent agents become equal to each other while varying (together) strongly with more kicks. Second, concentrating on dynamics whose Lyapunov exponent is small in magnitude, we found in Sec.~\ref{sec:sync} that the prevalence of low-entropy distributions $q(\varphi)$ permits accurate effective synchronization. 

With these results in hand, we outline one possible communication protocol between two remote agents. Before departing, the two agents must first agree on acceptable KLD and entropy thresholds. Then, each agent's $N$ oscillators are used to infer those quantities. For instance, by having each agent split their oscillator population into two, they may independently estimate the KLD and decide if their initial oscillator configurations have been forgotten based on whether the KLD lies below the agreed-upon KLD threshold. Then, between successive kicks, if the independently estimated entropy lies below the agreed-upon entropy threshold, communication should take place using their independently inferred fiducial phases (both rotating on the phase circle at the same rate), where the uncertainty is known by each agent to be $\sim e^S/\sqrt{N}$. If an unfavorable kick has occurred (resulting in a high entropy), each agent would independently be able to decide to cease communication until the next beneficial kick. Given the phenomena we have established in our paper, the independent decisions are guaranteed to be consistent between the two agents. Thus we establish effective synchronization, wherein the individual phase samples are not synchronized, but a collective phase variable that is constructed independently out from them\,---\,the fiducial phase\,---\,is. The extent to which the fiducial phases agree varies stochastically over time, but can independently be predicted by estimating the current entropy.

Below, we assess the potential impact of these results.  First, we note the narrow though significant scope of our explicit study.  We argue that this generalized synchronization should occur generically for oscillators subjected to non-synchronizing noise, despite possible degrading effects we ignored. 
We then discuss the power-law trade-off we observed between the desirable rapid convergence and the obtainable precision of the generalized synchronization. Finally, we consider how generalized synchronization might be relevant in cryptography, quantum-scale communication, and for understanding forms of cooperative behavior in living systems.

\subsection{Limitations}

Our demonstration of effective synchronization was made in a narrow context.  Among the various types of noise treated in the literature~\cite{TeramaePRL2004,YamamotoOE07}, we considered only impulsive noise whose effect could be described by a phase map~\cite{NakaoPRE2005,SongPRE2022}. Furthermore, the bulk of our study used a simple class of such maps\,---\,a cubic polynomials; other phase maps may not culminate in low-entropy distributions or have a meaningful fiducial phase. These limitations mean that our example is not immediately applicable to realistic conditions. Still, we may argue that the effect is somewhat general and robust.  The effective synchronization persists under continuous variation of the phase map $\psi(\varphi)$ through cubic and quintic phase maps and over a range of Lyapunov exponents $\Lambda$.  Further, we recover the effective phase $\varphi_\f$ using standard-precision computations and conventional sampling methods. This is not a delicate effect analogous to time-reversing a chaotic trajectory.   Moreover, all of our results in Sec.~\ref{sec:sync} show a common variation with the distribution entropy $S$, $\Delta \varphi_\f \sim e^S/\sqrt{N}$, under a range of underlying dynamics that gave rise to the distribution. Under general noise, one may define the Lyapunov exponent~\cite{TeramaePRL2004,NagaiPRE2009}, the distribution of $\varphi$ values $q(\varphi)$  at a given time, the notion of independence of this $q(\varphi)$ on initial conditions, and the determination of a fiducial phase $\varphi_\f$ from a given $q(\varphi)$.  Thus, the synchronization we found in our narrow context should plausibly have a counterpart for clocks subjected to more general forms of noise.

In addition, we did not address counter-effects that oppose synchronization, such as direct interactions~\cite{ZhouPRL02,MotterPRL20}, the presence of an additional degrading noise that acts independently on each oscillator~\cite{NakaoPRL2007,GoldobinPRE2005}, and inevitable differences between the two agents’ oscillators~\cite{GoldobinPRE2005}. Further, the two oscillators will in general experience random differences in arrival times of the kicks or other noise. These must all be included in order to reliably model, e.g., ecological systems, where spatial variations in the environment or inherent differences in the individuals are of interest. Nonetheless, we believe that achieving our strong-noise-induced effective synchronization in the presence of interactions or nonidentical perturbations is a realistic aim in many situations. Our belief is based in part on existing studies of conventional noise-induced synchronization ($\Lambda<0$), where the same degrading effects noted above have been studied and found manageable~\cite{GoldobinPRE2005,NakaoPRL2007}. Analogous manageable degradation should therefore be plausible for our effective synchronization ($\Lambda>0$). Explicitly, for $\Lambda<0$, it was shown that an additional, uncommon noise ultimately led to an asymptotic $q^{(k)}(\varphi)$ whose nonzero width is determined by the magnitude of the uncommon noise~\cite{GoldobinPRE2005,NakaoPRL2007}.\footnote{These references do not consider the effect of timing uncertainty explicitly, though Ref.~\cite{GoldobinPRE2005} did consider the effect of differing oscillator frequency, which we expect to be more damaging.} We expect a similar conclusion for $\Lambda>0$, in which the scaling $\sim e^S/N^{1/2}$ should worsen as a function of the degrading noise’s magnitude. Furthermore, the combination of interactions and shared noise was shown to still permit synchronization in the human brain~\cite{PangNEURON21} and in quantum circuits~\cite{BittnerJCP25}. These aspects will be explored in a future work.

Although our prescription for identifying $\varphi_\f$ is somewhat ungainly, the result is about as precise as one might hope.
The statistical fluctuations of $\Delta \varphi_\f$ appear consistent with an $N$-sample average of a Gaussian distribution whose variance is of order $N \langle\Delta \varphi_\f^2\rangle$, where $\langle\cdots\rangle$ denotes the average over samples.  The entropy of the $\Delta \varphi_\f$ distribution is thus of order $\ln(\langle\Delta \varphi_\f^2\rangle^{1/2}) + \log N$~\cite{Entropy2018lower}.  Now, the scaling shown in Fig.~\ref{fig:deviation_vs_entropy} amounts to saying $\ln(\langle\Delta \varphi_\f^2\rangle^{1/2})\simeq S - \log N$, \textit{i.e.}, $S$ is comparable to the entropy of $\Delta\varphi_\f$. Thus, a much more precise $\varphi_\f$ than the observed one would need to contain more information than the source distribution $q(\varphi)$ from which it was derived\,---\,a contradiction. Thus, we cannot expect other methods to yield qualitative improvements in precise synchronization compared to $\Delta\varphi_\f\sim e^S/\sqrt{N}$. Nevertheless, one can hope for more insightful and efficient calculation methods, which might or might not involve a fiducial phase. 

A clear practical limitation of our procedure was the laborious calculations it required.  Partly, this labor was due to our choice to demonstrate precise synchronization, \textit{i.e.}, small $\Delta \varphi_\f$.  This required using phase maps that produce low-entropy distributions (small positive $\Lambda$). This small $\Lambda>0$ implies slow convergence of $q(\varphi)$, as shown in Sec.~\ref{sec:conv}.  (For our example of Sec.~\ref{sec:sync}, the convergence number $K_\mathrm{m}$ was about $10^2$; knowledge of the last $\sim10^2$ kicks was needed in order to infer the converged $\varphi_\f$.) If instead we had been content to show synchronization to only a few bits of precision, we could have relaxed our requirement of small entropy and small $\Lambda>0$. Then, the large number of kicks needed to obtain a converged $\varphi_\f$ would be correspondingly reduced, thus reducing the required computation. We now consider further this tradeoff between computational labor and precision.

\subsection{Theoretical implications} \label{sec:theor_impl}

The central feature of our dynamics that makes effective synchronization possible is the property of finite-time convergence; the final distribution $q(\varphi)$ depends only on the recent history of waiting times. Although rigorous proofs, based on the spreading-of-phases argument, explain why asymptotic convergence is a possible outcome~\cite{BuzziCMP99,GuptaMRL13,FroylandNONLIN14,FroylandERG19}, the behavior of the convergence time had not been characterized before. Its predictability is central to making our scheme practically relevant in the first place.
Even more so, curiously, we found that the amount of history required had a simple dependence on $\Lambda$; the effective number of relevant kicks, $K_\mathrm{m}$, followed a power law. 

In general, when an iterated map becomes independent of its initial state, the characteristic kick number for convergence $K_\mathrm{m}$ has little relation to $\Lambda$. $\Lambda$ characterizes the separation rate of adjacent points while $K_\mathrm{m}$ depends on the merger of separated regions. Nonetheless, we empirically find a systematic variation of $K_\mathrm{m}$ with $\Lambda$ and, even more so, that $K_\mathrm{m}$ diverges as $\Lambda \to 0$. Characterizing this divergence further showed an apparent near-$3/2$ power-law dependence ranging across two orders of magnitude in $K_\mathrm{m}$. 
In Sec.~\ref{sec:conv}, we have given one qualitative argument for this interdependence, wherein $\Lambda>0$ implies that the phase map ``expands and fold'' neighboring phases, which in turn is responsible for the convergence. Without the former ($\Lambda\to0^+$), the latter should not occur ($K_\mathrm{m}\to\infty$).
We emphasize, however, that we have no evidence for such interdependence beyond the cubic map; understanding these scalings merits a future study.

As a corollary, we mention the close resemblance that noise-induced synchronization bears to the concept of random attractors~\cite{BOOK:RandAttract}. Just as deterministic systems described by ordinary differential equations may possess different types of attractors, random dynamical systems may admit attractors that depend on the noise realization. A specified noise realization determines the time-dependent basin of attraction and its region of support, which may differ between noise realizations and will keep evolving in time with subsequent noise. Our passage along the $\Lambda$ axis from evolution towards a (stochastic set of) synchronized single phase to a state characterized by a distribution of phases is reminiscent of known random dynamical systems at a bifurcation point. These bifurcating systems may shed light on our system, particularly our observed scaling of relaxation number of kicks $K_\mathrm{m}$ with the Lyapunov parameter $\Lambda$. Conversely, the effective synchronization demonstrated here may have analogs in these bifurcating systems.

\subsection{Practical implications}

We expect that the effective synchronization explored above may appear in a range of contexts. Here, we conjecture some of the possible utilities of common noise. For example, common noise adds to the possible ways that evolved systems such as biological organisms might exploit complex dynamics to generate subtle, adaptive behavior. Extending our strong-noise approach to account for interacting or stochastic clocks might be promising and relevant extensions of our work. They would allow testing whether noise-based encoding schemes might capture communication among, \textit{e.g.}, neurons such as central pattern generators~\cite{MarderCB01,YusteNAT05} and motion-sensitive neurons~\cite{BOOK:spikes1997,SteveninckSCI97}.

Further, our mechanism might offer promising implications, for example, in cryptographic contexts~\cite{GilpinPNAS18}. We posit that two remote agents may communicate amid chaos by utilizing the ambient noise as a time-dependent encryption key, potentially making it challenging for eavesdropping parties to interfere. Decoding the messages would require noting only the recent history of noise. 
 
Another potential application of our mechanism is to synchronization-based quantum communication systems~\cite{ArgyrisNAT05,ChoiMOBILE17,PljonkinCRYPT17}. With the growing recognition that noise-induced synchronization can occur also in qubit chains~\cite{LutzPRL22,LutzNC25}, we suggest that in the regime where the noise does not entangle the edge spins, they nonetheless should be partially correlated by virtue of statistical convergence onto low-entropy distributions. 

\section{Conclusion}

In conclusion, our effective synchronization mechanism amid strong noise shows an implicit organization, \textit{i.e.}, synchronization, hidden in apparent chaos. It permits two or more agents to generate a stream of shared information that facilitates synchronization and thereby enables cooperative activity. It seems worthwhile to explore such mechanisms further.

\begin{acknowledgments}

We benefited from discussions with Charlie Edelson, Efi Efrati, Gautam Reddy, Colin Scheibner, and Chuanyi Wang. We also thank Gil Ariel, Eli Barkai, Omer Granek,
Andrej Ko\v smrlj, Adilson Motter, Muhittin Mungan, Colin Scheibner, and Tony Song for helpful comments on the manuscript. B.S. acknowledges support from the Princeton Center for Theoretical Science. B.S. would like to thank the members of the James Frank Institute of the University of Chicago for their warm hospitality, where this research was initiated. 

\end{acknowledgments}

\appendix

\section{Example of a phase map}\label{sec:example_phase_map}

In Sec.~\ref{sec:prel}, we defined a ``phase map'' $\psi(\varphi)$ which encapsulates the effect of a kick on a particular nonlinear oscillator.  Here, we illustrate how this phase map is determined for a specific dynamical system and type of kick.

\begin{figure}
    \centering
    \includegraphics[width=0.99\linewidth]{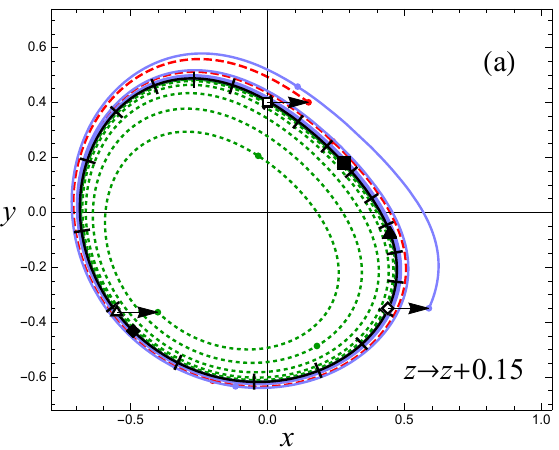}
    \includegraphics[width=0.99\linewidth]{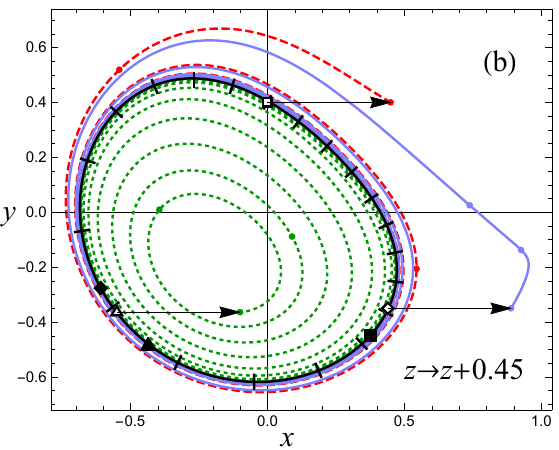}
    \caption{Trajectories from the dynamical system of Eq.~\eqref{eq:DW-SL}. The limit cycle (period $T=1.98$) is shown in full black oval. The black ticks along it are equispaced $\varphi\in[0,1)$ values, separated by $\Delta t/T=0.05$. ($\varphi=0$ is located at $0+0.4i$, overlapping with the empty square.) We show the resulting trajectories from kicking oscillators positioned at $\varphi=0$ (empty square), $\varphi\simeq0.35$ (empty triangle), and $\varphi\simeq0.6$ (empty diamond) by (a) $z\to z+0.15$ and (b) $z\to z+0.45$. The first three small points on each trajectory are the positions of the oscillators after $t=0,(1/2)T,T$. After a time $9T$, the oscillators reach the stable orbit with new phase values, $\psi\simeq-0.13$ (full square), $\psi\simeq0.73$ (full triangle), and $\varphi\simeq0.36$ (full diamond) in (a), and $\psi\simeq-0.44$ (full square), $\psi\simeq1.37$ (full triangle), and $\varphi\simeq-0.66$ (full diamond) in (b).}
    \label{fig:example}
\end{figure}

In this example, we construct the phase map for perturbations around the limit cycle of a modifed Stuart-Landau dynamics in the complex plane $z$. Here, $z\equiv x+iy$ evolves according to
\begin{subequations}
\label{eq:DW-SL}
\begin{align}
    \frac{dz}{dt}=\frac{1}{2}\bigl[&f(z+0.4+0.4i,5,2.75)\nonumber\\&+f(z-0.4-0.4i,-1,-4)\bigr],
\end{align}
where
\begin{equation}
    f(z,a,b)=(1+ai)z-(1+bi)|z|^2z.
\end{equation}
\end{subequations}
The conventional Stuart-Landau model~\cite{Stuart_1960,Watson_1960} exhibits either a stable fixed point or a harmonic limit cycle, characterized by a fixed radius and uniform angular velocity on the $z$ plane. Here, we aim to show that the phase-reduction procedure applies to oscillators with more complex limit cycles, particularly those which are not a circle in phase space nor have a uniform tangential velocity. The numbers in Eq.~\eqref{eq:DW-SL} were chosen such that a stable periodic orbit would be observed; see full black oval in Fig.~\ref{fig:example}. The period is $T\simeq1.98$.
We designate some point along the limit-cycle orbit as the phase origin, shown as an open square in Fig.~\ref{fig:example}. We define the phase position $\varphi$ of given point on the orbit as the time required to move from the phase origin to the given point, relative to the period $T$ (the black ticks along the black oval in Fig.~\ref{fig:example}). Thus, at the phase position $\varphi=1$, the given point has traversed the whole cycle and returned to the phase origin.

We now consider the effect of two particular kicks, defined to be a displacement of $z$ by an amount $0.15 + 0 i$ in Fig.~\ref{fig:example}(a) and $0.45 + 0 i$ in Fig.~\ref{fig:example}(b). We have chosen this displacement so that any point $\varphi$ on the limit cycle returns to it virtually completely in a time $9 T$ or less. Thus after any number of cycles greater than $9$, the final phase $\psi$ of the oscillator is constant. Displacements bigger than $\simeq0.46$ cause some of the initial phases to escape the periodic orbit's basin of attraction into that of the stable fixed point at $z\simeq1.5-0.56i$. In that sense, the latter displacement (b) is a very strong forcing; the former (a) is much milder.

\begin{figure}
    \centering
    \includegraphics[width=0.99\linewidth]{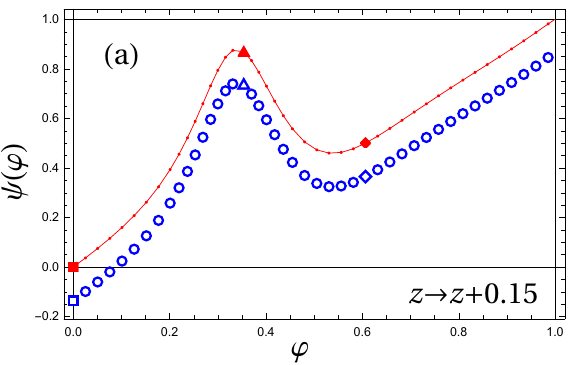}
    \includegraphics[width=0.99\linewidth]{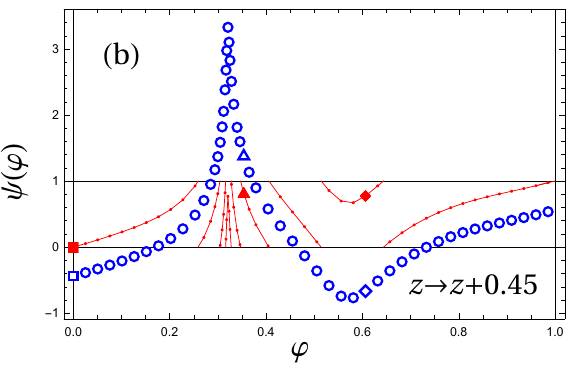}
    \caption{Phase lags from the dynamical system of Eq.~\eqref{eq:DW-SL}. Empty blue symbols are phase shifts as seen in Fig.~\ref{fig:example}. The square, diamond, and triangle symbols correspond to the three trajectories in Fig.~\ref{fig:example}.
    Since only the fractional parts of phases are significant, each symbol was shifted upwards or downwards by an integer so as to obtain a continuous dependence in $\varphi$.  In addition, this curve would be shifted vertically and horizontally by choosing a different point on the orbit as the phase origin. Red points show the corresponding phase map as we defined in the main text. Namely, the phase origin and integer-valued shifting were chosen so that $\psi(0) = 0$. Further, each $\psi(\varphi)$ value was shifted to lie between 0 and 1.
    }
    \label{fig:phasemap}
\end{figure}

If one kicks an oscillator that was at $\varphi=0$ (empty squares), it will then undergo the trajectories shown with the dashed red curves in Fig.~\ref{fig:example}. After $9T,10T,11T,\ldots$, we find, to good accuracy, that it reaches the pointx marked by solid squares, which is found to have a phase $\psi \simeq-0.13$ for the weak forcing (a) and $\psi \simeq -0.44$ for the strong forcing (b). (As before, $\psi$ is still defined $\mathrm{mod}\,1$. However, so to obtain the smooth collection of blue empty points in Fig.~\ref{fig:phasemap}, we determined integer offsets by counting the number of cycles completed during $9T$ relative to the $9$ cycles that an unperturbed oscillator would have completed during $9T$.) Likewise, kicking oscillators that are at $\varphi\simeq0.35$ (empty triangles) and $\varphi\simeq0.6$ (empty diamonds), through the dotted-orange and full-blue trajectories, they will, respectively, reach $\psi\simeq0.73$ and $\psi\simeq0.36$ for the weak forcing (a) and $\psi\simeq1.37$ and $\psi\simeq-0.66$ for the strong forcing (b), which are depicted with solid triangles and diamonds. We have picked $9T$ as our waiting time to guarantee that all oscillators have returns to the limit cycle within a small margin, so they can be once again uniquely described by a phase variable. We remind that were the three not kicked, after $9T$ they would have returned back to their original positions ($0$, $0.35$, and $0.6$), and that any phases that differ by an integer necessarily correspond to the same point on the orbit.

To determine the final phase $\psi$ for arbitrary initial phases $\varphi$, we repeat this process for many closely spaced initial $\varphi$'s. The result is the phase maps shown in Fig.~\ref{fig:phasemap}, which depend on the displacement (forcing magnitude), shown in Fig.~\ref{fig:phasemap}. The lags between the initial and final phases are shown with the empty blue symbols in Fig.~\ref{fig:example}. The squares, triangles, and diamonds correspond to the final phase $\psi$ of $\varphi=0$, $\varphi\simeq0.35$, and $\varphi\simeq0.6$, respectively, under each forcing. Constructing the phase map $\psi(\varphi)$ amounts to taking $\mathrm{mod}\,1$ of the resulting lags. Since the system is rotation invariant (as the next kick is also randomly timed according to a uniform distribution $[0,1)$), we shift all phases such that, arbitrarily, $\psi(\varphi=0)=0$. This allows the comparison with the cubic phase map of Eq.~\eqref{eq:phase_map}. Upon this shift and taking the moduli, we find the phase maps depicted with connected red points. Indeed the very strong forcing $z\to z+0.45$ leads to a very violent phase map (even having an apparent discontinuity in the spreading factor $d\psi/d\varphi$ at $\varphi\simeq0.32$). The mild forcing $z\to z+0.15$, which is the limit with which we are concerned in this work, has given rise to a well-behaved and smooth phase map with a single minimum and maximum. As we explained in Sec.~\ref{sec:prel}, broad minima and maxima are the ones facilitating the formation of sharply-peaked and thence low-entropy distributions. This suggests that our findings in Secs.~\ref{sec:conv} and~\ref{sec:sync} for the simplistic cubic phase map we chose should apply in `real-world' forcings as well.

\section{Binless estimation methods using samples}\label{sec:KLDentEst}

Here we explain how we estimate the phase distributions plotted in Fig.~\ref{fig:circ_q-phi}, and how we estimate the entropy and Kullback-Leibler divergence (KLD) throughout, using a finite number of samples $N$. 

At every timestep in our simulations, we generated $N$ phase samples per agent, $\{\varphi_{\A1}^{(k)},\ldots,\varphi_{\A N}^{(k)}\}$ and the same for $\B$. To draw Fig.~\ref{fig:circ_q-phi}, we used the following simplistic estimate:
\begin{subequations}
\begin{eqnarray}
    q_{\A n}^{(k)}&\equiv& q^{(k)}_\A\left(\varphi=\frac{\tilde\varphi^{(k)}_{\A n+1}+\tilde\varphi^{(k)}_{\A n}}{2}\right)\\&=&\frac{1}{N}\frac{1}{\tilde\varphi^{(k)}_{\A n+1}-\tilde\varphi^{(k)}_{\A n}}\label{eq:est_q-phi}
\end{eqnarray} 
\end{subequations}
(and the same for $q_\B^{(k)}$), where $\{\tilde\varphi^{(k)}_{\A n}\}$ are the instantaneous phase samples of agent $\A$, $\{\varphi^{(k)}_{\A n}\}$, arranged in increasing order on a ring ($\tilde\varphi^{(k)}_{\A n+1}>\tilde\varphi^{(k)}_{\A n}$, with a $\mathrm{mod}\,1$ constraint) and $1/N$ ensures normalization. In Sec.~\ref{sec:sync}, where a statistically significant estimate of $q^{(k)}(\varphi)$ was needed, we used, rather, the kernel-density estimate of Eq.~\eqref{eq:com_q-phi}.

Given these samples, we used established methods~\cite{VictorPRE2002,WangIEEE09} of estimating the entropy and KLD. These approximations can be conveniently expressed in terms of the nearest-neighbor distances between pairs of phases. Consider the $n$'th phase in ensemble $\A$ at iteration $k$ ($\varphi_{\A n}^{(k)}$); it has a unique closest distance to a phase in ensemble $\B$, $\varphi_{\B m}^{(k)}$ (where $m$ need not be equal to $n$). We denote the distance between these two as $\Delta_{\A n\leftarrow\B}^{(k)}$. (By the same logic, we also denote $\Delta_{\A n\leftarrow\A}^{(k)}$, $\Delta_{\B n\leftarrow\B}^{(k)}$, and $\Delta_{\B n\leftarrow\A}^{(k)}$.) Now, we estimate the entropy of agent $\A$'s distribution using~\cite{VictorPRE2002}
\begin{equation}
    S^{(k)}_\A=\frac{1}{N}\sum_{n=1}^N\ln\Delta^{(k)}_{\A n\leftarrow\A}+\ln(2N-2)+\gamma\label{eq:est_entropy}
\end{equation}
(and the same for $S_\B$), and the KLD of distribution $\A$ from $\B$ using~\cite{WangIEEE09}
\begin{equation}
    D^{(k)}(\A\Vert\B)=\frac{1}{N}\sum_{n=1}^N\ln\frac{\Delta_{\A n\leftarrow\B}^{(k)}}{\Delta_{\A n\leftarrow\A}^{(k)}}+\ln\frac{N}{N-1}\label{eq:est_KLD}
\end{equation}
(and the same for $D^{(k)}(\B\Vert\A)$), where $\gamma=0.577$ is the Euler-Mascheroni constant. See Figs.~\ref{fig:LposS&KLD} and~\ref{fig:LnegS&KLD} for a typical evolution of the entropy and KLD for $\Lambda=0.141$ and $\Lambda=-0.141$, respectively.

There are two technical limitations involving the above binless estimation methods. First is our limited sample size $N=500$, leading to an errorbar of $\pm0.2$ for entropy values (identified from samples of uniform distribution) and $\pm0.05$ for KLD values (identified from two sets of samples of the same Gaussian distribution). Second is the limited machine precision~\cite{SongPRE2022}; the phases are distinguishable only up to $\Delta_{\A n\leftarrow\B}^{(k)}>2^{-52}$ using the conventional $64\mathrm{bit}$ representation of real numbers. Each sample with $\Delta_{\A n\leftarrow\B}^{(k)}<10\times2^{-52}$ is ignored, and we replace $N\to N-1$.

The loss of samples to machine precision is a prevalent limitation. In the long realization ($k=10^4$) of Sec.~\ref{sec:sync}, $407$ and $439$ samples remained in the end for ensembles $\A$ and $\B$, respectively, which means that our estimates are still statistically meaningful. On the other hand, this implies that the low-entropy distributions (which are those that usually culminate in loss of samples) did not appear much; see Fig.~\ref{fig:deviation_vs_entropy}, where we see no distribution with $S<-21$. Upon continuing this simulation, there were no longer any distinguishable phase samples past $k=1.2\cdot10^4$, at which point the dynamics have attained entropies between $-34<S<-21$. Were lower-entropy distributions to appear, our results in Sec.~\ref{sec:sync} would improve further, as they are particularly useful for having a small discrepancy $\Delta\varphi_\f\sim e^S/\sqrt{N}$.

\section{Phase map iteration}\label{sec:est_q-phi}
\begin{figure}
    \centering
    \includegraphics[width=0.99\linewidth]{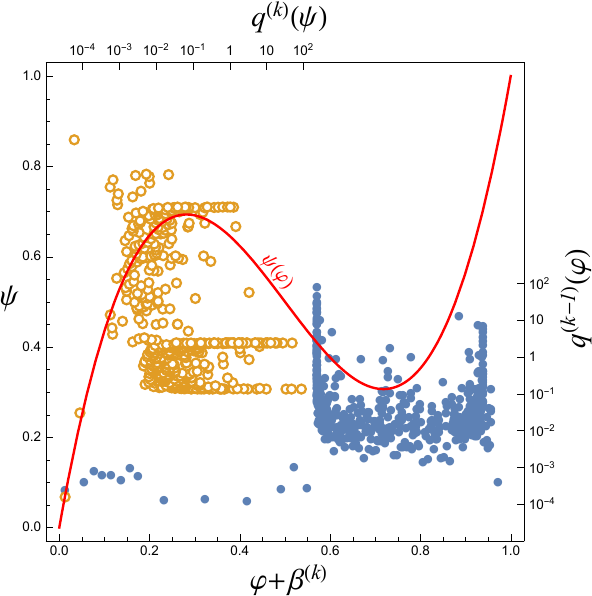}
    \caption{Two phase distributions $q^{(k)}(\varphi)$ before (blue full circles) and after (orange empty circles, with inflected axes) a kick, obtained for a typical realization of the cubic phase map (Eqs.~\eqref{eq:dynamics} and~\eqref{eq:phase_map}) with $A=9.32$ ($\Lambda=0.141$). The cubic phase map is drawn with a red line. One sees how the `sporadic' region of the initial distribution in the vicinity the phase map's extremum ($\varphi\simeq0.7$) is condensed into a prominent peak at $\psi\simeq0.3$. Conversely, the surroundings of the peak at $\varphi\simeq0.95$ are widened at $\psi=0.7$, while its height decreased from $\mathcal{O}(10)$ to $\mathcal{O}(1)$. The phase distributions are sampled from $N=500$ oscillators.}
    \label{fig:est_q-phi}
\end{figure}

In Fig.~\ref{fig:est_q-phi} we illustrate the changes in a set of initial phases $\{\varphi_n\}$ through an iteration of the phase map of Eq.~\eqref{eq:phase_map}.

\section{Dependence on stochastic realization}\label{sec:uncommon_noise}

\begin{figure}
    \centering
    {\large$\Lambda=-0.141$}
    \includegraphics[width=0.32\linewidth]{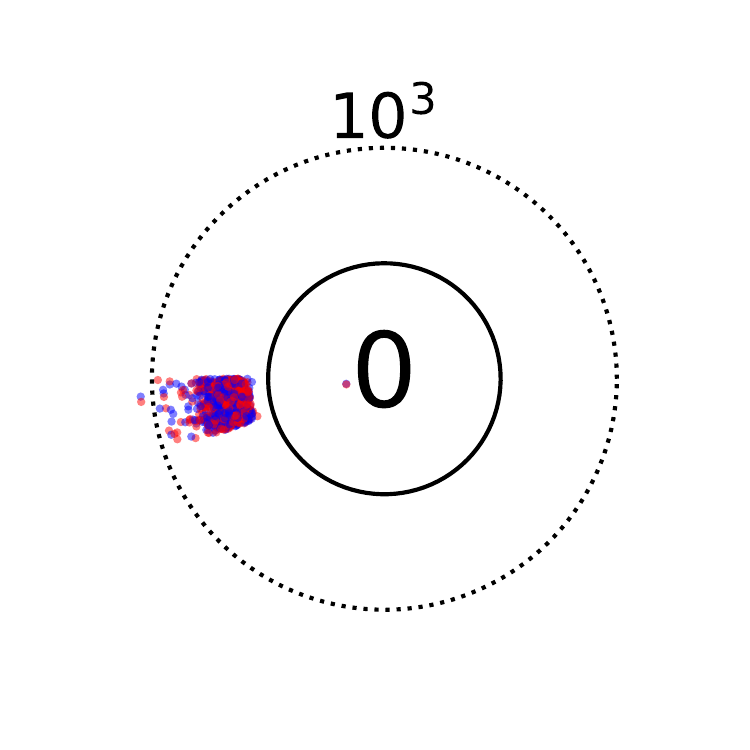}\includegraphics[width=0.32\linewidth]{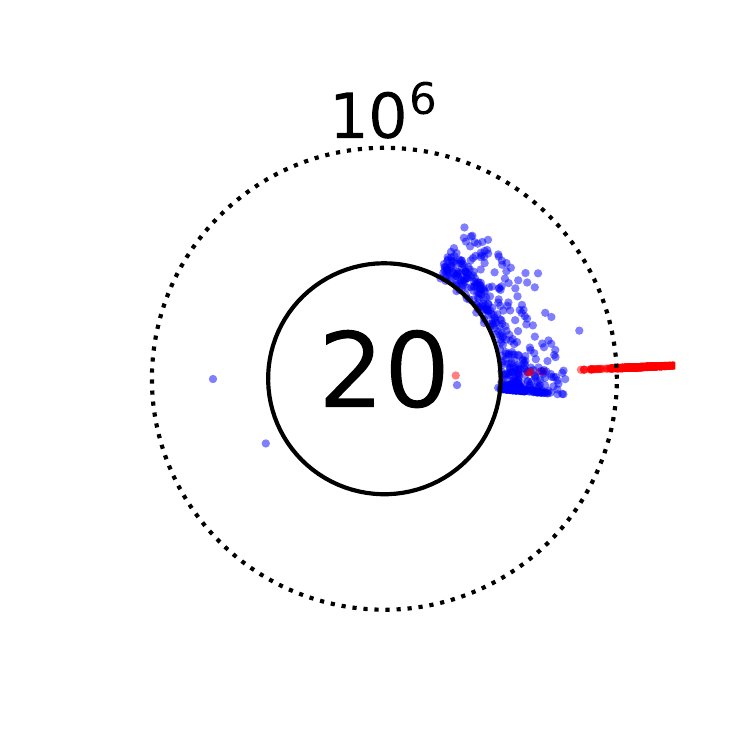}\includegraphics[width=0.32\linewidth]{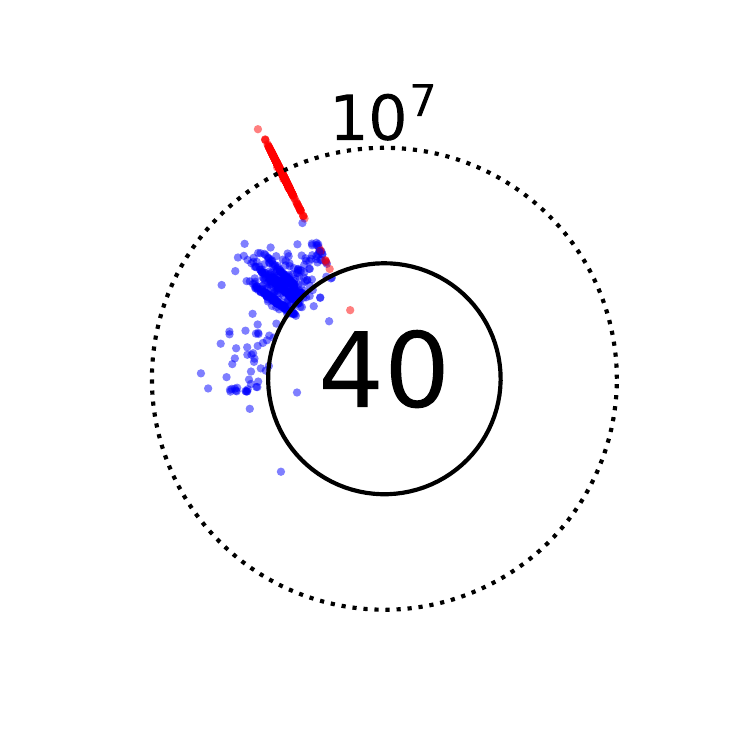}
    \includegraphics[width=0.32\linewidth]{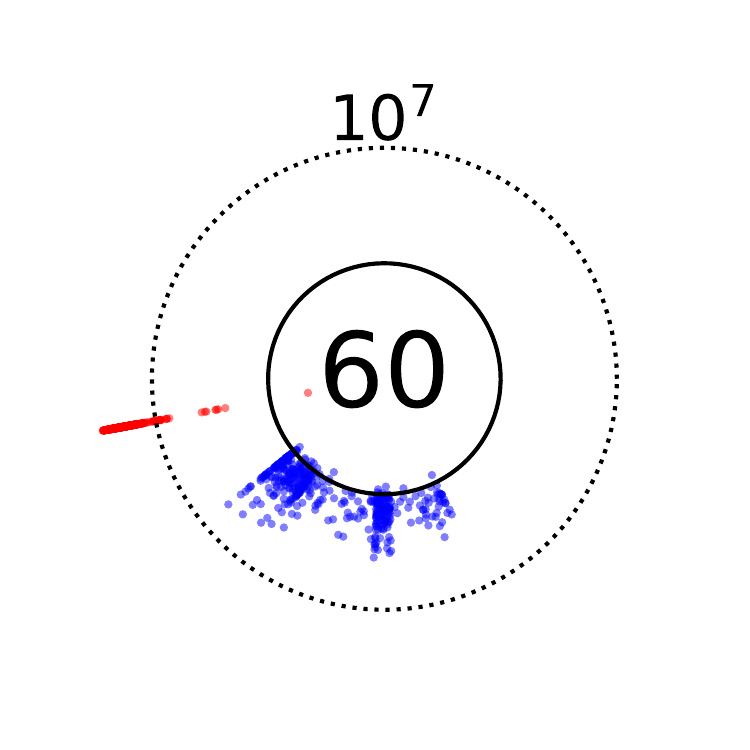}\includegraphics[width=0.32\linewidth]{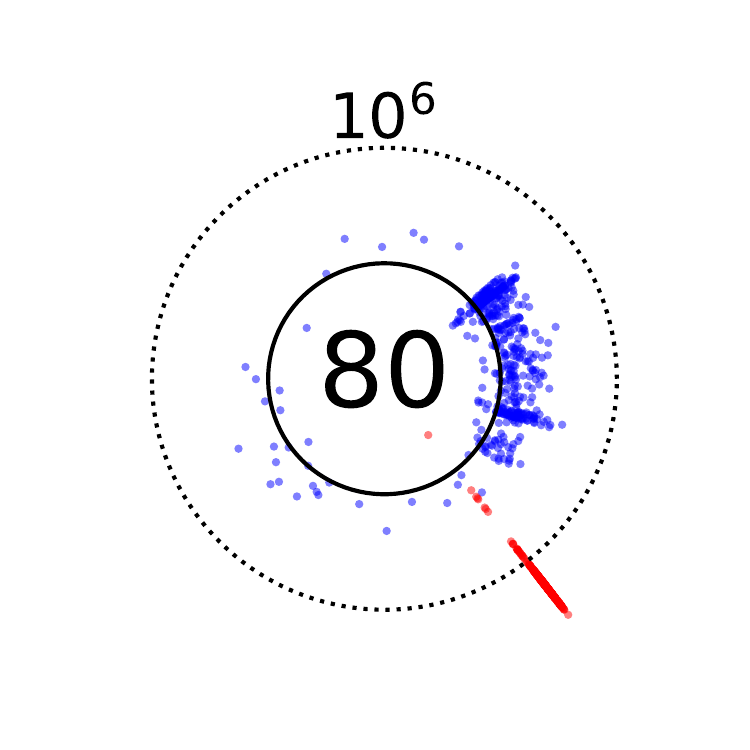}\includegraphics[width=0.32\linewidth]{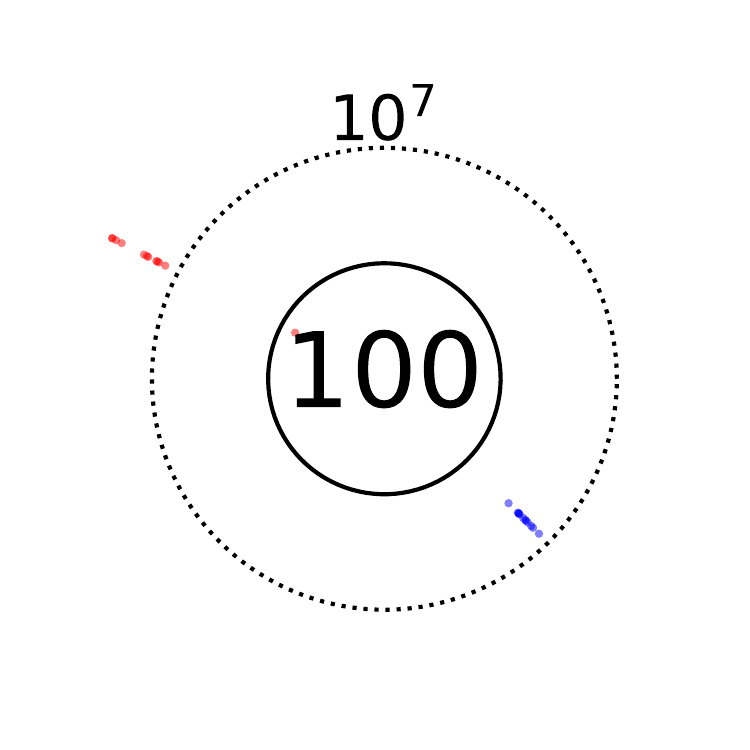}
    {\large$\Lambda=0.141$}
    \includegraphics[width=0.32\linewidth]{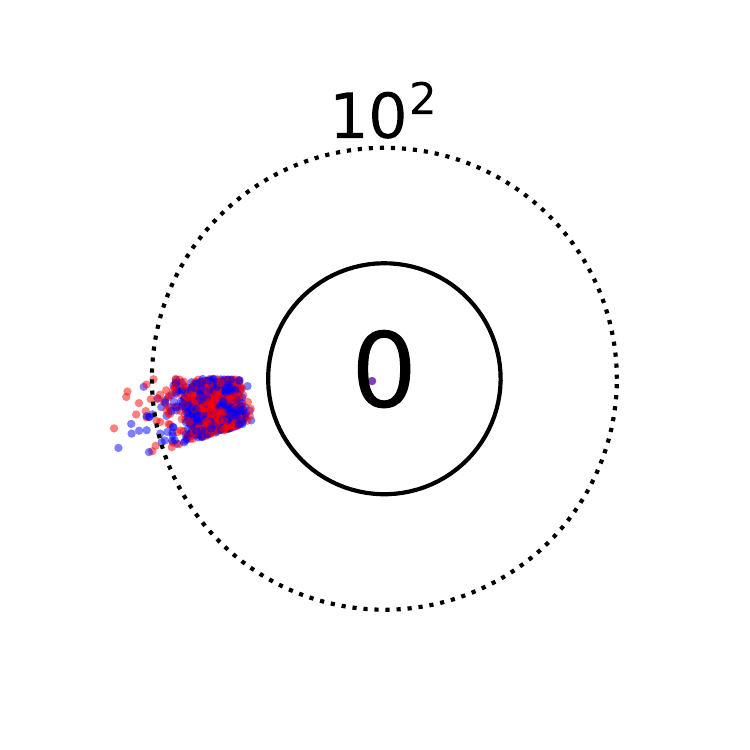}\includegraphics[width=0.32\linewidth]{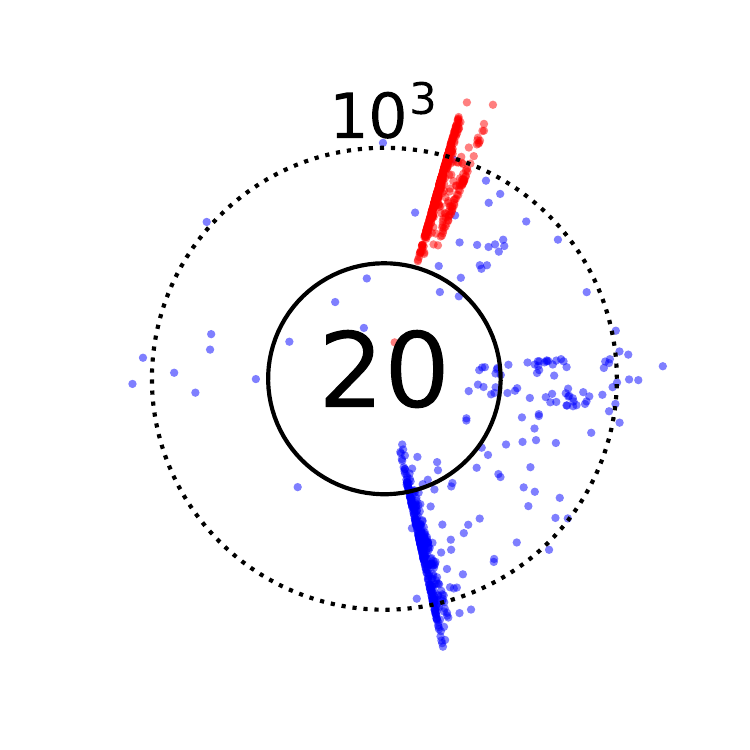}\includegraphics[width=0.32\linewidth]{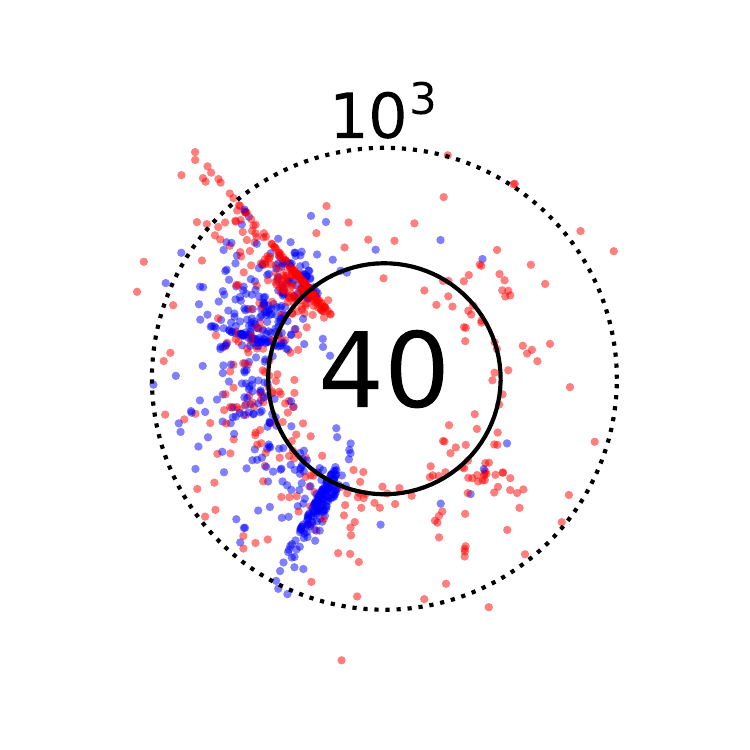}
    \includegraphics[width=0.32\linewidth]{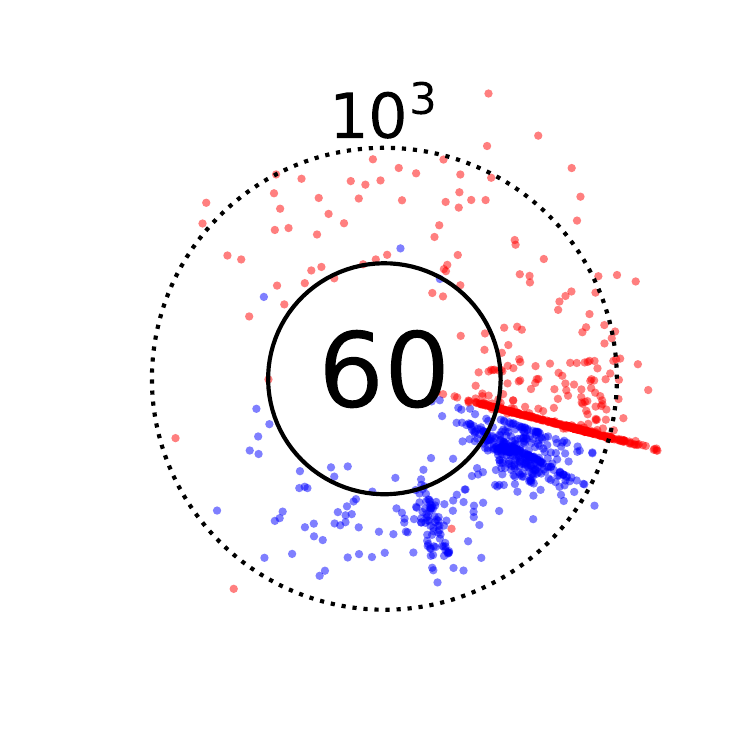}\includegraphics[width=0.32\linewidth]{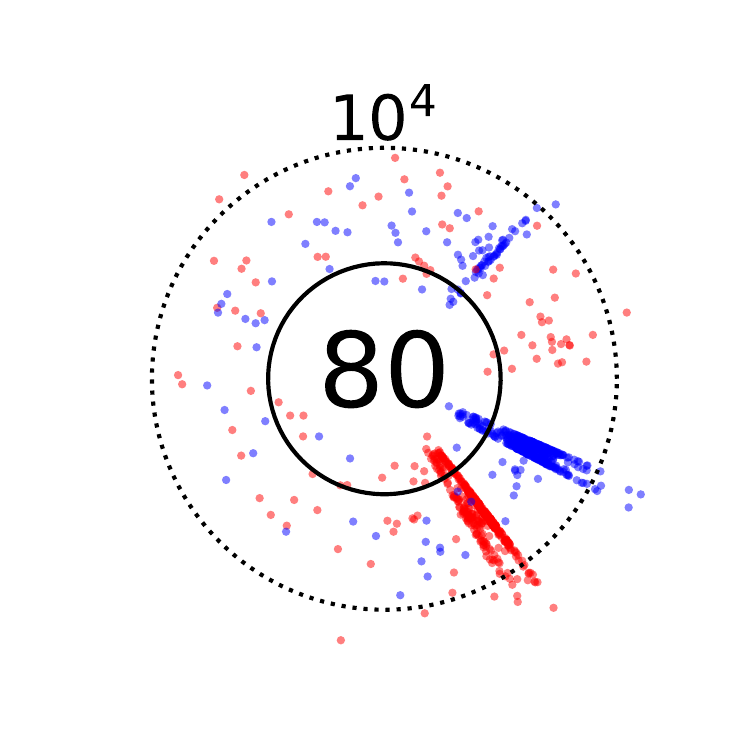}\includegraphics[width=0.32\linewidth]{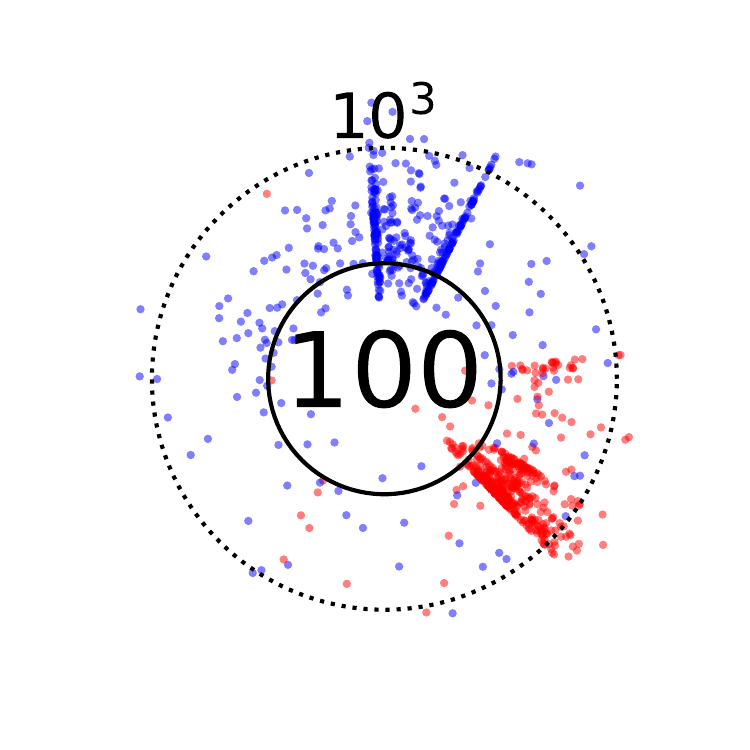}
    \caption{A sequence of two, initially-identical phase distributions $q^{(k)}_\A(\varphi)$ (red) and $q^{(k)}_\B(\varphi)$ (blue) subjected to different noises in a typical realization of Eqs.~\eqref{eq:dynamics} and~\eqref{eq:phase_map} with either $\Lambda=-0.141$ ($A=7.32$) or $\Lambda=0.141$ ($A=9.32$), as indicated. The initial distributions are uniform, $\varphi\sim\mathrm{U}[0.5,0.55)$. The distributions are drawn on a circle so the periodicity $\,\mathrm{mod}\,1$ of the phase circle is apparent. The radial axis shows the distributions' values $q(\varphi)$ on a log scale, where $q(\varphi)=1$ for the inner full circle and as indicated for the outer dotted circles. The kick numbers $k$ are shown inside the inner circle. Since the noise is different for each ensemble, $\Lambda<0$ produces different synchronized states, whereas $\Lambda>0$ produces erratic distributions that do not converge unto each other. The distributions are sampled using $N=500$ oscillators via the nearest-neighbor distances; see Eq.~\eqref{eq:est_q-phi}. The initial samples drawn for ensemble $\A$ are copied to ensemble $\B$, so the starting point is truly identical.}
    \label{fig:SMcirc}
\end{figure}

In Fig.~\ref{fig:circ_q-phi}, we show a sequence of distributions obtained for two agents under the influence of common noise. Despite the different initial conditions, for $\Lambda<0$ we observe synchronization, whereas for $\Lambda>0$ we observe convergence of ensembles. To emphasize that a common noise is a necessary condition, in Fig.~\ref{fig:SMcirc} we show two ensembles, starting from the same distribution, which are subjected to different noises. Evidently, since the information about the ambient noise is not shared, the distributions do not synchronize with $\Lambda<0$ and do not converge for $\Lambda>0$.

\section{Below-threshold synchronization}\label{sec:convergenceL<0}

\begin{figure}
    \centering
    \includegraphics[width=0.99\linewidth]{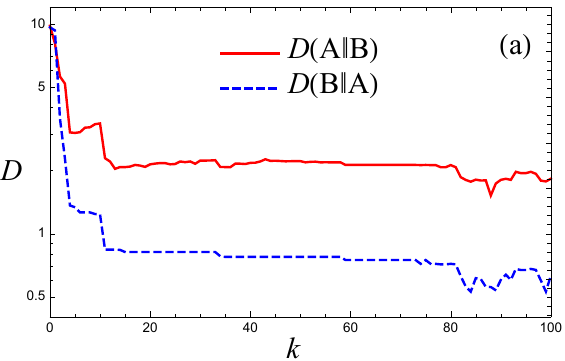}
    \includegraphics[width=0.99\linewidth]{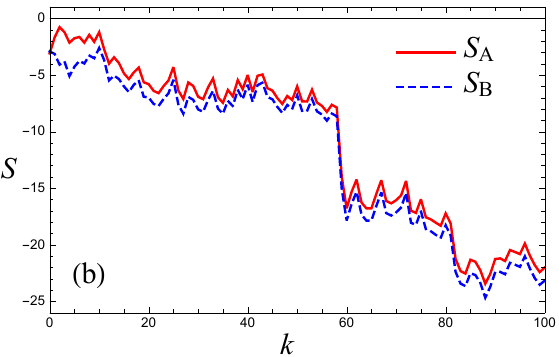}
    \caption{(a) Kullback-Leibler divergences (KLDs) $D(\mathrm{A}\Vert\mathrm{B})$ and $D(\mathrm{B}\Vert\mathrm{A})$ among two distributions and (b) entropies $S_\mathrm{A}$ and $S_\mathrm{B}$ of each distribution, as obtained for the waiting time realization $\{\beta^k\}$ of Fig.~\ref{fig:circ_q-phi} with $\Lambda=-0.141$. The number of phase samples is $N=500$. The initial distribution per ensemble is obtained with width $u=0.05$. The numerical errors in the estimation of the KLDs and entropies are $\Delta D=0.05$ and $\Delta S=0.2$, respectively.}
    \label{fig:LnegS&KLD}
\end{figure}

Here we revisit the discussion of Sec.~\ref{sec:conv} for a synchronizing phase map with $\Lambda < 0$.  That is, the agents' phases agree up to a gradually-diminishing uncertainty. For $\Lambda<0$, both distributions $q^{(k)}_\mathrm{A}(\varphi)$ and $q^{(k)}_\mathrm{B}(\varphi)$ are expected to become a sharp peak centered around the same phase value. To test this, we initiate both distributions to be concentrated uniformly into eccentric narrow sectors of angular width $u=0.05$, one centered at $\varphi=0.025$ and the other at $\varphi=0.525$. While indeed both entropies decay to $-\infty$ (see Fig.~\ref{fig:LnegS&KLD}(b)), both KLDs have reached a \emph{different nonzero} plateau (see Fig.~\ref{fig:LnegS&KLD}(a)). 

We can analyze this observation quantitatively. Since each $q_\mathrm{A}^{(k)}(\varphi)$ and $q_\mathrm{B}^{(k)}(\varphi)$ is very narrow, for a given $\beta^{(k)}$, at most a single $\hat\varphi$ contributes appreciably to a given $\varphi$ during a kick event. Namely, if there is a $\hat \varphi$ such that $\psi(\hat \varphi + \beta^{(k)})  = \varphi$ and $q_\mathrm{A}^{(k)}(\hat \varphi)$ is within its peak region, then it is given by
\begin{equation}
    q^{(k)}_\mathrm{A}(\varphi)\simeq q_\mathrm{A}^{(k-1)}(\hat{\varphi})\left|\left.\frac{d\psi}{d\varphi}\right
    |_{\hat{\varphi}+\beta^{(k)}}\right|^{-1},
\end{equation}
and the same for a narrow $q_\mathrm{B}^{(k)}$. Otherwise, the contribution from other $\varphi$'s is negligible. Thus, stochastic-timed forcings may only ``stretch'' or ``contract'' narrow unimodal distributions to the same extent. Since both $q^{(k)}_\mathrm{A}(\varphi)$ and $q^{(k)}_\mathrm{B}(\varphi)$ are centered around the same phase value, the same $\hat{\varphi}$ corresponds to each $\varphi$ in both distributions. Hence, upon a change of variables $d\varphi=(d\psi/d\varphi)|_{\hat{\varphi}+\beta^{(k)}}d\hat\varphi$, we find the KLD by definition
\begin{eqnarray}
    D^{(k)}(\mathrm{A}\Vert\mathrm{B})&=&\int_0^{1}d\hat\varphi q_\mathrm{A}^{(k-1)}(\hat{\varphi})\ln\frac{q_\mathrm{A}^{(k-1)}(\hat{\varphi})}{q_\mathrm{B}^{(k-1)}(\hat{\varphi})}\\&=&D^{(k-1)}(\mathrm{A}\Vert\mathrm{B}),
\end{eqnarray}
and similarly 
$D^{(k)}(\mathrm{B}\Vert\mathrm{A}) = D^{(k-1)}(\mathrm{B}\Vert\mathrm{A})$. Thus, indeed the KLD is capable of remaining constant for two infinitely narrow distributions, as they either ``stretch'' or ``contract'' \emph{together} to the same extent. Similar reasoning explains why, while the entropies decay to $-\infty$, the difference among them for these two narrow distributions remains constant for low entropy values in Fig.~\ref{fig:LnegS&KLD}(b).

Thus, for $\Lambda<0$, it is possible to interpret the asymptotically-nonzero KLDs to mean that agent $\mathrm{B}$'s ``guess'' ($q_\mathrm{B}^{(k)}(\varphi)$) is not in perfect agreement with agent $\mathrm{A}$'s actual distribution ($q_\mathrm{A}^{(k)}(\varphi)$). This is misleading in our context, as efficient communication may still be established in the $\Lambda<0$ scenario. To the contrary, the distributions $q_\mathrm{A}^{(k)}(\varphi)$ and $q_\mathrm{B}^{(k)}(\varphi)$ described here permit $\varphi_\mathrm{f}$ to be transmitted to arbitrary precision. This nonzero divergence simply stems from $q_\mathrm{A}^{(k)}(\varphi)$ and $q_\mathrm{B}^{(k)}(\varphi)$ not necessarily attaining the same (almost-zero) width.

From similar consideriations, observe the $25\lesssim k\lesssim 65$ regime in Fig.~\ref{fig:LposS&KLD}(a), where the dynamics of $\Lambda>0$ is shown. As seen in Fig.~\ref{fig:circ_q-phi}, $\Lambda=0.141$ panels, indeed the distributions are very narrow during these steps (which also manifests in a low entropy in Fig.~\ref{fig:LposS&KLD}(b)). This regime mimics the $\Lambda<0$ case at late times, where the narrow distributions change width together and barely mix, which manifests in a constant KLD in Fig.~\ref{fig:LposS&KLD}(a). As we argued here, this nonzero KLD need not imply that these sharp distributions are not useful for communication, so long as they share means. However, since $\Lambda>0$, the distributions are bound to get sufficiently wide eventually, continue the mixing, and thus the KLD resumes its decay.

\section{Instantaneous fiducial phase}\label{sec:instantaneous}   

\begin{figure}
    \centering
    \includegraphics[width=0.99\linewidth]{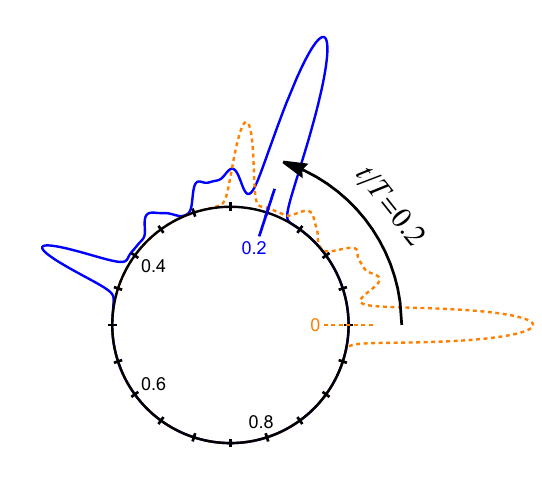}
    \caption{Relation between the fiducial phase position $\varphi_\mathrm{f}$ and the instantaneous fiducial phase $\varphi_\mathrm{f}+0.2T$ at a later time $t=0.2T$, as measured by a given agent. Orange dotted curve represents the distribution $q(\varphi)$ for some particular noise history.  The highest peak of this distribution is then $\varphi_\mathrm{f}$, and is marked with a dashed line.  At time $t=0$ this is the actual phase distribution.  At a time $t = 0.2 T$ later, every phase point, including the agent's own oscilllator's phase, has advanced through $0.2$ cycles. The probability distribution $q(\varphi-0.2T)$ and fiducial phase $\varphi_\mathrm{f}+0.2T$ at that moment is shown by the blue line. }
    \label{fig:rot_q-phi}
\end{figure}

Figure~\ref{fig:rot_q-phi} illustrates a phase probability distribution defined in Sec.~\ref{sec:prel}.  It pictures a time interval long after the most recent kick, but before the next kick.  It shows how the phase distribution $q(\varphi)$ and the fiducial phase $\varphi_\mathrm{f}$ move in time.

\section{Identifying a fiducial phase}\label{sec:fiducial}

To obtain $\varphi_\f$ we must find a smooth estimate of $q(\varphi)$ from the sampled values $\{\varphi_1,\ldots,\varphi_N\}$. For this purpose, we use the methodology of kernel-density estimation~\cite{bookSIGMA86,ParkSIGMA90,SheatherSIGMA91}. It produces the desired smooth distribution, $Q(\varphi,\sigma)$, by convolving a kernel function with the sampled phase positions $\{\varphi_n\}$,
\begin{equation}
     Q(\varphi;\sigma)=\frac1N\sum_{n=1}^NG(\varphi - \varphi_n;\sigma) ,  \label{eq:com_q-phi}
\end{equation}
The kernel function, $G(\varphi;\sigma)$, is parametrized by some nonzero smoothing window, or bandwidth, $\sigma$. Here, we choose the wrapped-Gaussian function,
\begin{equation}
    G(\varphi;\sigma)=\sum_{j=-\infty}^\infty\frac1{\sqrt{2\pi\sigma^2}}\exp\left[-\frac{(\varphi-j)^2}{2\sigma^2}\right].
\end{equation}
Given the estimated distribution $Q(\varphi, \sigma)$, we may determine its global maximum $\varphi_\f$ according to Sec.~\ref{sec:sync}.

For every $k$, based on the $N$ samples that each agent has, we compute the distributions $Q^{(k)}_\A(\varphi;e^{S_\A^{(k)}})$ and $Q^{(k)}_\B(\varphi;e^{S_\B^{(k)}})$ from Eq.~\eqref{eq:com_q-phi} on lattices of spacings $e^{S_\A^{(k)}}/10^3$ and $e^{S_\B^{(k)}}/10^3$, respectively. Using Eq.~\eqref{eq:fiducial}, we find $\varphi^{(k)}_{\f,\A}$ and $\varphi^{(k)}_{\f,\B}$. There are more involved discretization-free techniques to find extrema of kernel-density estimators~\cite{FukunagaSIGMAMAX75,ChengSIGMAMAX95,ComaniciuSIGMAMAX02}. Since $\varphi^{(k)}_{\f,\A}$ and $\varphi^{(k)}_{\f,\B}$ will only differ by about $e^S/\sqrt{N}$, with $N=500$, using a discretization $e^S/10^3$ should not impact the results.

\section{Effect of the bandwidth} \label{sec:smooth_q-phi}

\begin{figure}
    \centering
    \includegraphics[width=0.99\linewidth]{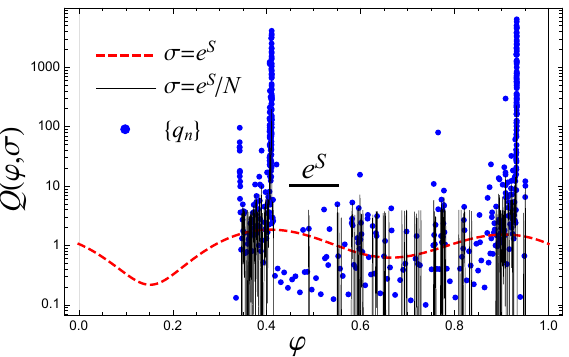}
    \caption{Illustration of the kernel-density estimation method. Here we estimate the distribution from $N=500$ phase samples obtained at $k=100$ by agent $\mathrm{B}$ during the realization of Fig.~\ref{fig:circ_q-phi} for $\Lambda=0.141$. The density estimate of Eq.~\eqref{eq:est_q-phi} is shown in points. The kernel-density estimate, $Q(\varphi;\sigma)$, is carried out with either $\sigma=e^S$ (the bandwidth we use throughout Sec.~\ref{sec:sync}; dashed red curve) or $\sigma=e^S/N$ (the alternative extreme; thin solid black curve). $S=-2.29$ for this distribution using Eq.~\eqref{eq:est_entropy}. A scale bar of $e^S$ is shown for reference with a thick black horizontal line. Clearly, the result of the latter bandwidth is not statistically significant, as it is comparable to the nearest-sample distances. The analysis in Sec.~\ref{sec:proof} is carried out with the former bandwidth. }
    \label{fig:smooth_q-phi}
\end{figure}

In Sec.~\ref{sec:sync}, we describe the estimation of a smooth probability function $Q(\varphi, \sigma)$ from a set of samples $\{\varphi_n\}$. In Fig.~\ref{fig:smooth_q-phi} we show the effect of a different bandwidth $\sigma$ using sampled $\{\varphi_n\}$'s.

\section{fiducial phase differences between agents}\label{sec:sigma_vary}

In Sec.~\ref{sec:sync}, we compared the fiducial phases obtained by two independent agents.  Their  oscillators were subjected to the same phase map $\psi(\varphi)$ and the same kick timings $\{\beta^{(k)}\}$. Since each agent used a finite set of $N=500$ samples to obtain their $\varphi_\mathrm{f}^{(k)}$ after the $k$'th kick, their fiducial phases may not fully coincide when they sample an identical distribution. Thus, the agents obtain different values for $\varphi_\mathrm{f}^{(k)}$. Figure~\ref{fig:deviation_vs_entropySM} shows the obtained differences $|\Delta\varphi_\mathrm{f}|$ as a scatter plot against the common distribution's entropy, using two different bandwidths, (a) $\sigma=e^S$ and (b) $\sigma=e^S/N$. We interpret these results in the next section.

\section{Rare misidentifications of fiducial phase}\label{sec:misidentification}

\begin{figure}
    \centering
    \includegraphics[width=0.99\linewidth]{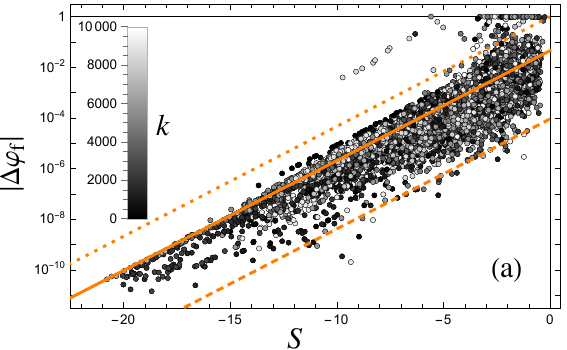}
    \includegraphics[width=0.99\linewidth]{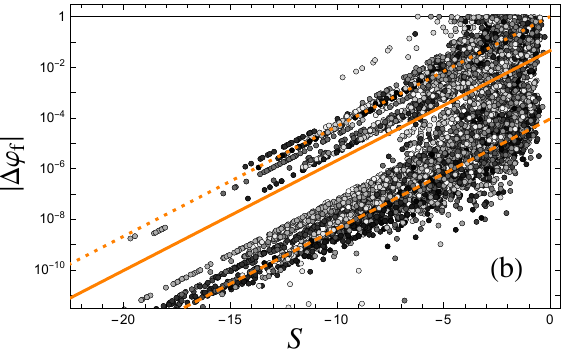}
    \caption{A scatter plot of deviations in the fiducial phases of each party $\Delta\varphi_\mathrm{f}^{(k)}=\varphi_{\mathrm{f},\mathrm{A}}^{(k)}-\varphi_{\mathrm{f},\mathrm{B}}^{(k)}$ against the entropy of their distributions $S^{(k)}=(S^{(k)}_\mathrm{A}+S^{(k)}_\mathrm{B})/2$, using bandwidths (a) $\sigma=e^S$ (used in Fig.~\ref{fig:deviation_vs_entropy}) and (b) $\sigma=e^S/N$. Each $(S,|\Delta\varphi_\mathrm{f}|)$ point is computed given a pair of distributions $Q_\mathrm{A}^{(k)}$ and $Q_\mathrm{B}^{(k)}$ obtained after kicks (with $\Lambda=0.141$) $k=1,\ldots10^4$. For completeness, the points' shading encodes the kick at which it was obtained, as indicated. The orange dotted, solid, and dashed lines are, respectively, $|\Delta\varphi_\mathrm{f}|=e^{S}$, $|\Delta\varphi_\mathrm{f}|=e^{S}/\sqrt{N}$, and $|\Delta\varphi_\mathrm{f}|=e^{S}/N^{3/2}$.}
    \label{fig:deviation_vs_entropySM}
\end{figure}

We point out that out of $10^4$ data-points for $|\Delta\varphi_\mathrm{f}|$ used in Sec.~\ref{sec:sync}, $88$ points lie above the $|\Delta\varphi_\mathrm{f}|=e^S$ line ($0.9\%$), typically even a few orders of magnitude above it for high-entropy distributions; see Fig.~\ref{fig:deviation_vs_entropySM}(a). On examination, these points either corresponded to distributions with entropy $S\sim0$ or proved to be misidentifications, in which the $Q(\varphi;e^S)$'s had two peaks of nearly equal height and the agents made opposite determinations of the global maximum, as a result of the finite samples used to estimate $Q(\varphi;e^S)$. Among these occurrences is the distribution of Fig.~\ref{fig:smooth_q-phi}, where each agent chose a different peak out of the two at $0.41$ and $0.93$. 

Based on this finding, we suggest that the effect of these large $\Delta \varphi_\mathrm{f}$'s can be much reduced.  Each agent is able to determine the uncertainty of its highest and second highest peak.  If the two heights are statistically indistinguishable, the agent cannot obtain an unambiguous $\varphi_\mathrm{f}$ and would not be able infer the other agent's effective phase at that kick.  Had this been done in our analysis, the large $|\Delta \varphi_\mathrm{f}|$'s could have been reduced for distributions whose entropies are not approaching $0$.  We anticipate that any remaining large $\varphi_\mathrm{f}$'s could be made insignificant.

These misidentifications were an important factor in our choice of smoothing width $\sigma=e^S$.  Indeed, when we used $\sigma=e^{S}/N$ (rather than $\sigma = e^S$ of Sec.~\ref{sec:proof}) as a bandwidth, there were $791$ misidentified points ($8\%$), which have occurred with distributions of entropies as a low as $S=-14$; see Fig.~\ref{fig:deviation_vs_entropySM}(b). In fact, the large concentration of points surrounding $|\Delta\varphi_\mathrm{f}|=e^S$ suggests that the bandwidth is too small, and not enough samples are involved in estimating the peak center, raising the error from $|\Delta\varphi_\mathrm{f}|\sim e^S/\sqrt{N}$ to just the overall ``width'' of a given distribution $|\Delta\varphi_\mathrm{f}|\sim e^S$. At the same time, in the cases where the identification of the peak was precise (that is, the highest peak clearly dominates over the rest and encapsulated $\sim N$ samples), this smaller bandwidth allowed for an appropriately smaller discrepancy, $|\Delta\varphi_\mathrm{f}|\sim (e^S/N)/\sqrt{N}$. Thus, as we expected, smaller bandwidths facilitate smaller deviations, which comes at a cost of more frequent, finite-sample-size-related misidentifications of the highest peak.

%\bibliography{nonlin_osc_BIB}

\end{document}